\documentclass[aps,prd,nofootinbib,twocolumn,superscriptaddress,preprintnumbers,letterpaper, longbibliography]{revtex4-1}

\usepackage{amssymb}
\usepackage{amsmath}
\usepackage{epsfig}
\usepackage[colorlinks=true
,urlcolor=blue
,anchorcolor=blue
,citecolor=blue
,filecolor=blue
,linkcolor=red
,menucolor=blue
,linktocpage=true
,pdfproducer=medialab
,pdfa=true
]{hyperref}
\usepackage{comment}
\usepackage{color}
\usepackage{cleveref}
\usepackage{multirow}
\usepackage{capt-of}
\usepackage{siunitx}
\DeclareSIUnit\electronvolt{e\kern-.05em V}

\usepackage{graphicx}
\usepackage{gensymb}
\usepackage[caption=false]{subfig}
\usepackage{slashed}
\usepackage{tabularx}
\usepackage{blindtext}
\usepackage[utf8]{inputenc}

\def\polarc{horizontal }
\def\polars{vertical }

\def\polarcly{horizontally }

\def\Ec{E_0^\rightarrow}
\def\Ep{E_+^\uparrow}
\def\Em{E_-^\uparrow}
\def\Epm{E_\pm^\uparrow}
\def\Ein{E_{0,\text{in}}^\rightarrow}
\def\rc{r_\rightarrow}
\def\rs{r_\uparrow}
\def\tc{t_\rightarrow}
\def\ts{t_\uparrow}

\def\Ref{Ref.~}

\begin{document}

\preprint{MIT-CTP/5048}

\title{
    Searching for Axion Dark Matter with Birefringent Cavities
}

\author{Hongwan Liu}
\email{hongwan@mit.edu}
\affiliation{Center for Theoretical Physics, Massachusetts Institute of Technology, Cambridge, MA 02139, U.S.A.}

\author{Brodi D. Elwood}
\email{bdelwood@mit.edu}
\affiliation{Massachusetts Institute of Technology, Cambridge, MA 02139, U.S.A.}

\author{Matthew Evans}
\email{m3v4n5@mit.edu}
\affiliation{Massachusetts Institute of Technology, Cambridge, MA 02139, U.S.A.}

\author{Jesse Thaler}
\email{jthaler@mit.edu}
\affiliation{Center for Theoretical Physics, Massachusetts Institute of Technology, Cambridge, MA 02139, U.S.A.}

\begin{abstract} 
Axion-like particles are a broad class of dark matter candidates which are expected to behave as a coherent, classical field with a weak coupling to photons. 
Research into the detectability of these particles with laser interferometers has recently revealed a number of promising experimental designs. Inspired by these ideas, we propose the Axion Detection with Birefringent Cavities (ADBC) experiment, a new axion interferometry concept using a cavity that exhibits birefringence between its two, linearly polarized laser eigenmodes. This experimental concept overcomes several limitations of the designs currently in the literature, and can be practically realized in the form of a simple bowtie cavity with tunable mirror angles. Our design thereby increases the sensitivity to the axion-photon coupling over a wide range of axion masses. 

\end{abstract}

\maketitle
\textit{Introduction}. The QCD axion is a well-motivated solution to the strong $CP$ problem which also provides a natural dark matter candidate~\cite{Peccei:1977hh,Peccei:1977ur,Weinberg:1977ma,Wilczek:1977pj,Preskill:1982cy,Abbott:1982af,Dine:1982ah}, particularly in the \SI{}{\micro\eV} to \SI{10}{\milli\eV} range~\cite{Graham:2015ouw}. More generally, however, dark matter could be made up of light, pseudoscalar axion-like particles (ALPs). These are generically predicted in string theory, and can have masses much less than a \SI{}{\micro\eV}~\cite{Jaeckel:2010ni,Svrcek:2006yi,Arvanitaki:2009fg,Acharya:2010zx,Cicoli:2012sz}. For brevity, we will often use the word ``axion'' to refer to any such ALP dark matter candidate.  

ALP dark matter would behave as a coherent, classical field $a$, and may couple weakly to photons through the following interaction term: 
\begin{alignat}{1}
    \mathcal{L} \supset -\frac{1}{4} g_{a\gamma\gamma} a F_{\mu\nu} \tilde{F}^{\mu\nu}.
    \label{eqn:axion_EM_interaction}
\end{alignat}
Such an interaction has motivated a large range of experimental searches for ALPs converting into photons (and vice versa) in the presence of a static magnetic field~\cite{Sikivie:1983ip,Wilczek:1987mv}. This includes proposed and ongoing searches for unexpected modifications to the amplitude, phase, or polarization of propagating light~\cite{Cameron:1993mr,Tam:2011kw,DellaValle:2015xxa}, light-shining-through-walls experiments~\cite{Chou:2007zzc,Robilliard:2007bq,Ehret:2010mh,Betz:2013dza,Ballou:2014myz,Ballou:2015cka}, the conversion of axion oscillations into electromagnetic waves~\cite{Horns:2012jf,Chaudhuri:2014dla,Kahn:2016aff,TheMADMAXWorkingGroup:2016hpc,Foster:2017hbq,Chaudhuri:2018rqn,Du:2018uak,Baryakhtar:2018doz,Ouellet:2018beu,Ouellet:2019tlz}, and axion helioscopes~\cite{Anastassopoulos:2017ftl,Armengaud:2014gea}. Among these, past and ongoing experiments~\cite{Cameron:1993mr,DellaValle:2015xxa,Chou:2007zzc,Robilliard:2007bq,Ehret:2010mh,Betz:2013dza,Ballou:2014myz,Ballou:2015cka,Horns:2012jf,Du:2018uak,Ouellet:2018beu,Anastassopoulos:2017ftl} have steadily improved constraints on $g_{a\gamma\gamma}$, and have even begun probing couplings that are relevant to the QCD axion~\cite{Du:2018uak}. 

Laser interferometry without a strong, static magnetic field has also been shown to be an effective way of searching for axions~\cite{Melissinos:2008vn,DeRocco:2018jwe,Obata:2018vvr}. The interaction term in Eq.~(\ref{eqn:axion_EM_interaction}) causes a difference in phase velocity between left- and right-handed circularly polarized light, and an appropriately designed high-finesse Fabry-Perot cavity can be used to accumulate the resulting phase difference. These studies have shown how to exploit the exquisite sensitivity of interferometry to small phase differences to obtain new limits on low mass axions.

Despite their ingenuity, these designs face two key limitations. First, they are limited by the non-ideal behavior of optical elements. The introduction of a beam splitter~\cite{Melissinos:2008vn} or quarter-wave plates~\cite{DeRocco:2018jwe} inside a cavity leads to losses and imperfect phase shifts between polarization modes that accumulate with each pass of laser light in the cavity. The authors of \Ref\cite{Obata:2018vvr} attempted to overcome this difficulty by using a bowtie cavity; however, circularly polarized light is not in general a bowtie eigenmode, as reflection off any surface at a nonzero angle of incidence does not preserve circular polarization. These difficulties would have to be addressed for an actual realization of these proposals.

Second, and more importantly, these proposed experiments rely on the coherent build-up of the phase difference over the entire light storage time in the cavity. The sensitivity of these experiments starts to deteriorate once the axion oscillation period becomes comparable to the storage time, i.e.\ when $m_a \ell \sim 1/\mathcal{F}$, where $\ell$ is the length of the cavity, $\mathcal{F}$ is the finesse, and $m_a$ is the mass of the axion. Increasing $\mathcal{F}$ therefore restricts the experimental sensitivity to lower axion masses, even though a large value of $\mathcal{F}$ is desirable to maximize a possible axion signal. 

In this paper, we propose a new axion interferometry experimental design that simultaneously overcomes both of these limitations. The presence of ALP dark matter results in a rotation of \polarcly polarized laser light propagating with frequency $\omega_0$ in a cavity, causing a small, \polars polarization to develop in the frequency sidebands $\omega_0 \pm m_a$. We exploit the fact that oblique reflection generally results in a phase difference between different linear polarizations to design a cavity that is resonant at $\omega_0$ in the \polarc (carrier) polarization, and $\omega_0 \pm m_a$ in the \polars (signal) polarization. The signal sidebands can then be detected using conventional interferometry techniques. Our design is sensitive to axion masses $m_a \lesssim 1/\ell$ independent of the finesse of the cavity, significantly improving the reach in $m_a$ without compromising on the strength of the axion signal. All of this can be achieved by a simple, practical cavity design requiring only that light reflect off multiple mirrors at oblique angles. 

\vspace{0.2cm}
\textit{Axions and Light Polarization}. Consider two orthogonal, circular polarizations of a laser beam (denoted by $\circlearrowright$ and $\circlearrowleft$) propagating with frequency $\omega_0$ and wavenumber $k_0$ in the presence of an axion field $a(t) = a_0 \cos(m_a t - k_a z)$, starting at some time $t_0$. The axion momentum is $k_a = m_a v$, where $v \sim 10^{-3}$ is the typical dark matter velocity at the Earth. We will only be interested in $m_a \ell \lesssim 1$, so that $k_a \ell \ll 1$, allowing us to neglect spatial gradients in the axion field.

The interaction term in Eq.~(\ref{eqn:axion_EM_interaction}) leads to the following dispersion relation for the two polarizations:
\begin{alignat}{1}
    - \omega_0^2 + k_0^2 = \pm k_0 g_{a\gamma\gamma} \frac{\partial a}{\partial t} \,.
\end{alignat}
After some time $t_{\circlearrowright, \circlearrowleft}$, each polarization travels a distance $\ell$, given by
\begin{alignat}{1}
    \ell = \int_{t_0}^{t_0 + t_{\circlearrowright, \circlearrowleft}} \left[ 1 \mp \frac{G}{\omega_0} \cos(m_a t) \right] \, dt \,,
    \label{eqn:polarization_travel_time}
\end{alignat}
where $G \equiv g_{a\gamma\gamma} \sqrt{2 \rho_\text{DM}}/2$, and $\rho_\text{DM} = m_a^2 a_0^2/2$ is the local density of dark matter. Equating the result from each polarization on the right-hand side of Eq.~(\ref{eqn:polarization_travel_time}), and working out the phase difference between the two polarizations $\Delta \alpha \equiv \omega_0 (t_{\circlearrowright} - t_{\circlearrowleft})$ to first order in $G/m_a$, we obtain  
\begin{alignat}{1}
    \Delta \alpha \simeq \frac{iG}{m_a} \left[ e^{i m_a t_0} \left(e^{i m_a \ell} - 1\right) + e^{-i m_a t_0} \left(1 - e^{-i m_a \ell} \right)\right]\, .
    \label{eqn:phase_difference}
\end{alignat}
Eq.~(\ref{eqn:phase_difference}) makes it clear that the axion field takes a carrier wave with frequency $\omega_0$ and generates signal sidebands with frequencies $\omega_0 \pm m_a$. 

This phase difference between circular polarizations is equivalent to a rotation of linearly polarized light. Writing the complex electric field in each circular polarization as a vector $(E^\circlearrowright,\, E^\circlearrowleft)$ and keeping track of the relative phase difference only, the translation matrix over a distance $\ell$ can be expressed as $\text{diag}(e^{i \Delta \alpha/2},\, e^{-i \Delta \alpha/2})$. The circular polarizations are related to the linear polarizations via $E^{\circlearrowright,\circlearrowleft} = E^\rightarrow \mp i E^\uparrow$, so that in the linear polarization basis $(E_\uparrow, E_\rightarrow)$, the matrix for translation is 
\begin{alignat}{1}
    P = \begin{pmatrix}
        \cos \frac{\Delta \alpha}{2} & - \sin \frac{\Delta \alpha}{2} \\
        \sin \frac{\Delta \alpha}{2} & \cos \frac{\Delta \alpha}{2} 
    \end{pmatrix} \simeq \begin{pmatrix}
        1 & - \frac{\Delta \alpha}{2} \\
        \frac{\Delta \alpha}{2} & 1 
    \end{pmatrix} \,.
    \label{eqn:2x2_translation_matrix}
\end{alignat}
\vspace{0.2cm}

\begin{figure}
    \centering
    \includegraphics[scale=0.34]{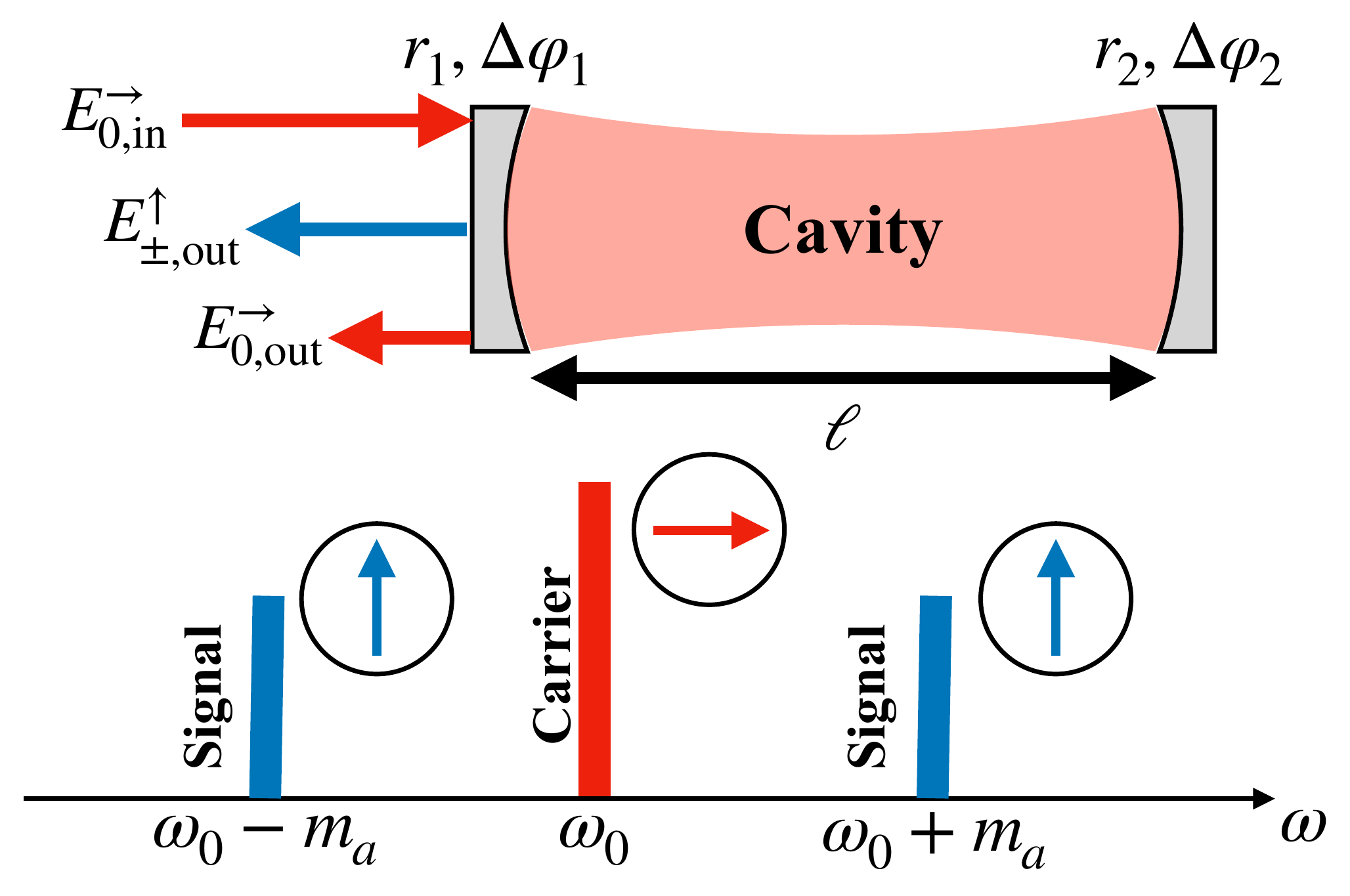}
    \caption{Summary of axion interferometry. A \polarcly polarized laser fed into a cavity with reflectivities $r_{1,2}$ and a relative phase shift between horizontally and vertically polarized light $\Delta \varphi_{1,2}$ at each end leads to the generation of frequency sidebands in the \polars polarization.}
    \label{fig:axion_interferometry_cartoon}
\end{figure}

\textit{Axion Interferometry.} The basic principle of axion interferometry is summarized in Fig.~\ref{fig:axion_interferometry_cartoon}. A carrier wave with electric field $\Ec$ in the \polarc polarization is injected into a cavity that is tuned to be resonant in the \polarc polarization at the laser carrier frequency $\omega_0$. As the field propagates in the presence of axions, signal sidebands in the \polars polarization are generated, with frequencies $\omega_0 \pm m_a$. The amplitude of the sidebands can be enhanced using an appropriately tuned high-finesse Fabry-Perot cavity. At each end of the cavity, a reflection occurs at a mirror with some real reflectivity coefficient, and a phase difference $\Delta \varphi_{1,2}$ between horizontally and vertically polarized light.

In order to distinguish between the two sidebands, we split the \polars signal polarization into its two frequency components by writing the electric field in the cavity as the complex column vector $\mathbf{E}_\text{cav} = (\Em,\, \Ec,\, \Ep)$. The subscripts indicate that the components have different frequencies $(\omega_0 - m_a,\, \omega_0,\, \omega_0 + m_a)$, respectively. The transfer matrix for translation in our 3-component notation follows from Eq.~(\ref{eqn:2x2_translation_matrix}):
\begin{alignat}{1}
    T \simeq \begin{pmatrix}
        e^{-i m_a \ell} & \frac{iG}{2m_a}(e^{-i m_a \ell} - 1) & 0 \\
        0 & 1 & 0 \\
        0 & \frac{iG}{2m_a} (1 - e^{i m_a \ell}) & e^{i m_a \ell}
    \end{pmatrix}.
    \label{eqn:3x3_translation_matrix}
\end{alignat}

For reflection at either end, we can write the transfer matrix as $R_{1,2} = r_{1,2} \, \, \text{diag} \left(e^{i \Delta \varphi_{1,2}},\,\, 1,\,\, e^{i \Delta \varphi_{1,2}} \right)$. The signal field in the cavity is then given by the solution to the following equation~\cite{Maggiore:1900zz}:
\begin{alignat}{1}
    \mathbf{E}_\text{cav} = t_1\, \mathbf{E}_0 + R_1 T R_2 T \, \mathbf{E}_\text{cav} \,,
    \label{eqn:E_cav}
\end{alignat}
where $\mathbf{E}_0 = (0, \Ein, 0)$ is the electric field of the laser fed into the cavity, and $t_X = \sqrt{1 - r_X^2}$ is the field amplitude transmission coefficient. 

Axion interferometry shares many parallels with conventional microwave cavity experiments like ADMX~\cite{Du:2018uak}. In both, the axion converts a frequency mode pumped to a large energy density (a DC magnetic field in microwave cavities, $\omega_0$ in our set-up) into another mode related to the original by $m_a$ (a standing electromagnetic mode of frequency $m_a$ in microwave cavities, and the signal sidebands $\omega_0 \pm m_a$ in our set-up). This conversion between electromagnetic modes is a generic property of ALPs coupled to photons through Eq.~(\ref{eqn:axion_EM_interaction}), as studied more generally in \Ref\cite{Goryachev:2018vjt}. 

The parallel extends to the power stored in both cavities. In the laser cavity, the power stored in the signal sidebands within the cavity is $P_\pm \propto |\Epm|^2 w^2$, where $w$ is the laser beam width. Solving Eq.~(\ref{eqn:E_cav}) gives $P_\pm \sim g_{a\gamma\gamma}^2 (\rho_\text{DM}/m_a) E_0^{\rightarrow 2} V Q_\pm$, where $V \sim w^2 \ell$ is the volume encompassed by the beam, and $Q_\pm$ is a quantity dependent on the geometry of the cavity, and is analogous to the quality factor for microwave cavities. This reproduces the scaling of the signal power produced in ADMX, with $E_0^\rightarrow = B_0^\rightarrow$ for the laser. 

\vspace{0.2cm}
\textit{Birefringent Cavities}. We now turn our attention to the importance of the phase difference between horizontally and vertically polarized light in the cavity, $\Delta \varphi_{1,2}$. Birefringence in a cavity has been used by the PVLAS experiment~\cite{DellaValle:2015xxa} to look for axion-induced changes in the polarization of a propagating laser beam in the presence of a large, static magnetic field due to the Primakoff effect. In contrast, our set-up relies on light transitioning between polarizations due to the absorption or emission of axions. 

In Ref.~\cite{Melissinos:2008vn}, a single beam passes through a polarizing beam splitter so that each beam propagates over a different path length along two different, perpendicular arms, effectively introducing birefringence between the two polarizations. However, a cavity with two perpendicular arms and a beam splitter at its center is unlikely to have a high finesse. More recent work has always ensured that $\Delta \varphi_1 = \Delta \varphi_2 \simeq 0$ either by using quarter-wave plates in front of mirrors with near-zero transmission~\cite{DeRocco:2018jwe}, or by performing two reflections at each end of the cavity, separated by an optical path length that is much shorter than the cavity length~\cite{Obata:2018vvr}. The signal generated by the axion builds constructively as long as the axion field value does not change significantly during the storage time, i.e.\ $m_a \mathcal{F} \ell \ll 1$. Once $m_a \sim 1/(\mathcal{F} \ell)$, the cavity loses sensitivity to the axion signal. 

An equivalent way of understanding the loss of sensitivity without birefringence is to observe that setting $\Delta \varphi_{1,2} = 0$ means that light in both polarizations is resonant at the laser frequency $\omega_0$. The full-width half-maximum of the cavity transmission band is $\delta \lambda \sim 1/(\mathcal{F} \ell)$, and so we must have $m_a \ll \delta \lambda \sim 1/(\mathcal{F} \ell)$ in order for the signal sidebands (produced by axion-driven polarization modulation) to lie within the transmission band. 

Now consider the birefringent case where $\Delta \varphi_{1,2} = \Delta \varphi \neq 0$ (we take $r_2 = 1$ in the following discussion for simplicity). When the resonance condition in the signal polarization $m_a \ell = |\Delta \varphi|$ is met, the signal polarization builds constructively in the cavity. With a phase shift of $\Delta \varphi = \pm \pi/2$, the cavity is resonant at $m_a = \pi/(2\ell)$, the maximum mass reach, for the sidebands $\omega_0 \pm m_a$. Axion masses up to this maximum value can be scanned by increasing $\Delta \varphi$ in steps from 0 to $\pi/2$. Since a larger finesse $\mathcal{F}$ is desirable for producing a large signal field, this represents a significant improvement in axion mass reach without affecting the sensitivity in coupling. Although higher frequency resonances exist for each choice of $\Delta \varphi$, the axion field value at these higher frequencies oscillates more than once over the cavity length $\ell$, suppressing the sensitivity by $\text{sinc}(m_a \ell)$~\cite{Maggiore:1900zz}. 
\begin{figure}
    \centering
    \includegraphics[scale=0.31]{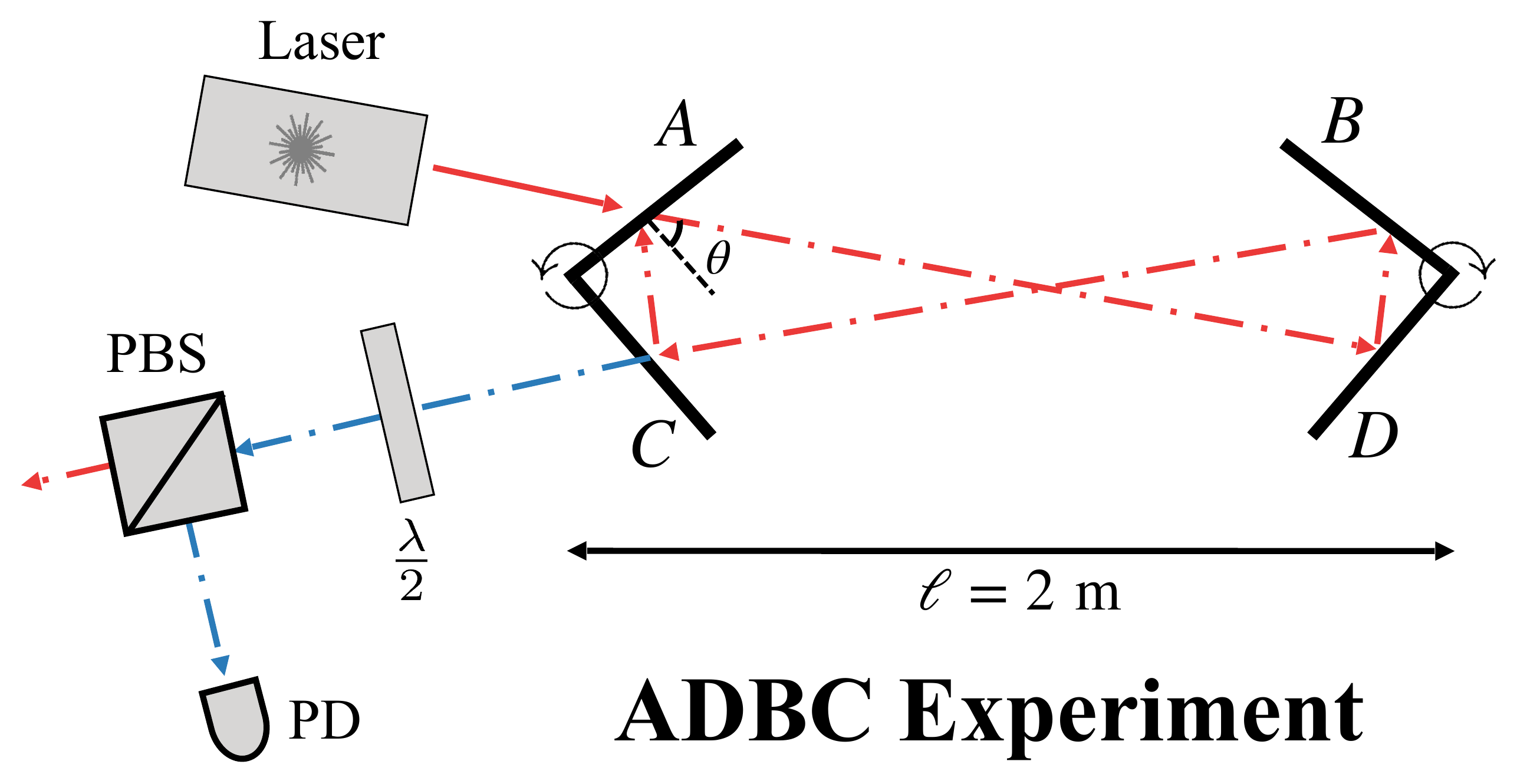}
    \caption{Schematic of the ADBC experiment. The red optical path is that of the input and cavity, while the blue optical path is read-out. The beam enters at $A$ and is read out after $C$. Two sets of mirrors $A$, $C$ and $B$, $D$ can be rotated to change the angle of incidence $\theta$ while roughly maintaining cavity alignment and length. To produce an electrical signal, the leakage fields from mirror $C$ pass through a half-wave plate ($\lambda/2$) before reflecting off a polarizing beam splitter (PBS) and arriving at a photodetector (PD).
\label{fig:design}}
\end{figure}

\vspace{0.2cm}
\textit{Experimental Set-up}. 
\label{experimentalsetup}
Fig.~\ref{fig:design} gives a schematic of the proposed Axion Detection with Birefringent Cavities (ADBC) experiment, featuring a practical cavity design with the necessary birefringence. The Fresnel equations~\cite{hecht2016optics} show that orthogonal, linear polarizations reflecting off a dielectric surface at an oblique angle of incidence $\theta$ in general develop a relative phase shift $\Delta \varphi$. By rotating the mirrors to adjust the angles of incidence, we can thus tune the cavity birefringence to make the axion-induced, vertically polarized sidebands resonant in our cavity at $m_a \ell = |\Delta \varphi|$.

The proposed design consists of two sets of two mirrors spaced \SI{2}{m} apart, with each set acting as a retroreflector that can pivot independently.  The angle between the mirrors in a set should be fixed at slightly less than $90\degree$, so that the angles of incidence are roughly $\theta$ and $90\degree - \theta$.  This allows us to vary the angle of incidence while roughly maintaining optical path-length and cavity alignment. The short dimension of the cavity (e.g.\ $\ell_{DB}$) is of order \SI{10}{cm}. One set, $A$ and $C$, will be taken as our input and output ports respectively, so that the optical path goes in the order $ADBC$. 

The Fresnel equations show that the reflectivity of the horizontal polarization will be lower than the vertical. Placing the carrier in the horizontal polarization (lower finesse) therefore reduces the accumulation of experimental noise in the cavity, while simultaneously placing the signal in the vertical polarization (higher finesse) leads to a larger signal-to-noise ratio (SNR). 

To prevent appreciable leakage of the carrier from the cavity, the cavity should be optimally coupled, meaning the transmissivity of $A$ in the carrier polarization, $\tc^A$, must be matched to the total losses in the cavity. This would almost entirely eliminate any reflection off $A$.  To allow a significant signal field to be read out, we also need $\ts^C$ to be larger than the total losses from the other mirrors. However, the Fresnel equations force $\tc > \ts$, and as a result, cavity loss for the carrier will be dominated by $\tc^C$, leaving us with $\tc^A \simeq \tc^C$.
To maintain high finesse in the signal and carrier, all other transmissivities should be smaller than the cavity optical loss.

To maximize the axion mass reach, the mirrors should cover as much of the range $0 \leq \Delta \varphi \leq \pi/2$ as possible. $\Delta \varphi$ increases with more oblique angles of incidence, but large optical surfaces are required near grazing incidence.

\vspace{0.2cm}
\textit{Experimental Sensitivity}. 
\label{experimentalsensitivity} The sensitivity of ADBC to $g_{a\gamma\gamma}$ is ultimately dependent on the finesse of the cavity $\mathcal{F}_\uparrow$ and $\mathcal{F}_\rightarrow$ in each polarization, and on $t_{\rightarrow,\uparrow}^C$, for which we will use benchmark values of $\mathcal{F}_\uparrow = 2.25 \times 10^5$, $\mathcal{F}_\rightarrow = 2700$, $t_\uparrow^C = 0.0037$, and $t_\rightarrow^C = 0.030$ (recall that $t_X$ is the \textit{amplitude} transmission coefficient). These finesse values are typical of mirrors used in the LIGO cavity and other experimental studies~\cite{Isogai:2013ic}. The reach in axion mass is determined by $\Delta \varphi$, which in turn depends on the mirror properties. We find that a range of  $0 < \Delta \varphi \lesssim \pi/5$ can typically be probed over a $6\degree$ range in the angle of incidence $\theta$, with $\theta \lesssim 65\degree$. Over this small range of angles, the finesse of the cavity does not vary significantly in either polarization, and we have adopted the smallest values in this range for simplicity. 

The signal field inside the cavity can be found by solving this cavity's equivalent of Eq.~(\ref{eqn:E_cav}) for $\mathbf{E}_\text{cav}$. For simplicity, we neglect the translation matrix for the short legs (i.e.\ $\ell_{DB}$ and $\ell_{CA}$), and take the same matrix $R$ for both sets of mirrors.
The reflection matrix has the form $R = \text{diag} (\rs e^{i \Delta \varphi}, \, \, \rc,  \,\, \rs e^{i \Delta \varphi})$, with $\rc^2$ and $\rs^2$ being the product of the reflectivities of all four cavity mirrors. These quantities are related to the finesse by $\mathcal{F}_{\uparrow, \rightarrow} \simeq \pi/(1 - r_{\uparrow,\rightarrow}^2)$.

\begin{figure}
    \centering
    \includegraphics[scale=0.58]{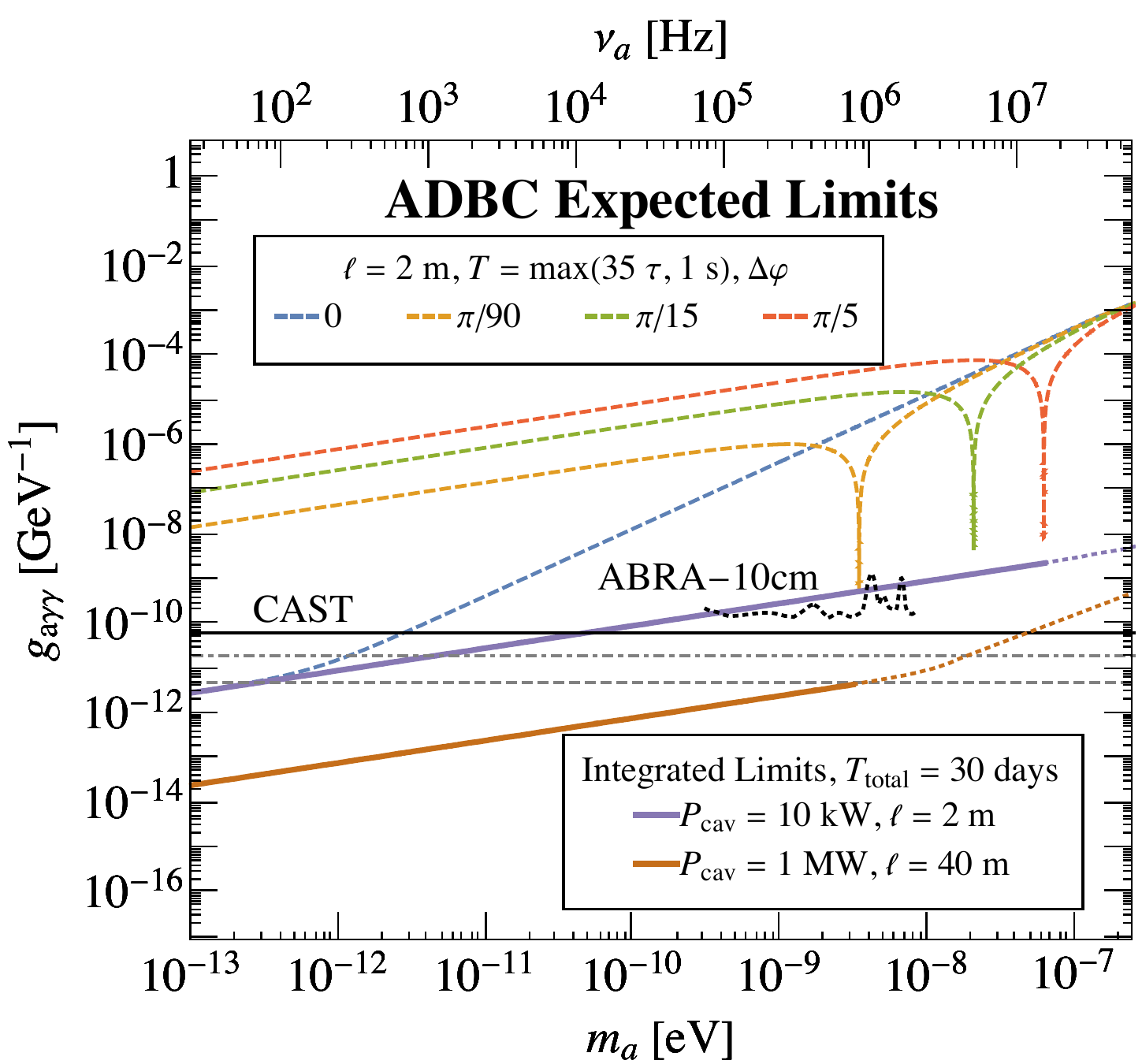}\caption{
    Expected ADBC limits on the axion coupling $g_{a\gamma\gamma}$. The limits for a phase shift of 0 (blue, dashed), $\pi/90$ (orange, dashed), $\pi/15$ (green, dashed) and $\pi/5$ (red, dashed) are shown. The integrated limits obtained by scanning through a phase shift of 0 to $\pi/5$ are shown for a 2 m cavity with 10 kW laser power in the cavity (purple) and for a 40 m cavity with 1 MW laser power in the cavity (brown), with a total integration time of 30 days. The envelope of the reach can be extended in both the 2 m (purple, dotted) and 40 m cavities (brown, dotted) to higher axion masses if the optics were improved to scan up to $\Delta \varphi = \pi/2$. Limits from CAST~\cite{Anastassopoulos:2017ftl} (black) and ABRACADABRA-10cm~\cite{Ouellet:2018beu} (black, dotted), together with projected limits from ALPS-II~\cite{Bahre:2013ywa} (gray, dot-dashed) and IAXO~\cite{Irastorza:1567109} (gray, dashed) are shown for comparison. These results have no dependence on the polarizer angle $\varepsilon$, since the SNR in Eq.~(\ref{eqn:SNR}) is independent of $\varepsilon$.}
    \label{fig:ADBC_reach}
\end{figure}
The signal sidebands emerging from the cavity are read out using a heterodyne detection scheme. The carrier and signal are passed through a half-wave plate with its fast axis rotated by a small angle $\varepsilon$ away from vertical, after which a polarizing beam splitter (PBS) is used to isolate the vertical polarization for readout by a photodetector. This mixes a small amount of the DC (carrier) electric field component, which is proportional to $\varepsilon$, into the AC (signal sideband) electric field component. The DC field acts as a local oscillator with frequency $\omega_0$, while the sidebands oscillate at a frequency $\omega_0 \pm m_a$; this produces a power modulation at the frequency corresponding to the axion mass. If the phase difference is tuned so that $\Delta \varphi = m_a \ell$, the cavity is resonant in the vertical (signal) polarization at a frequency sideband $\omega_0 - m_a$, giving an output AC power at the heterodyne readout of
\begin{alignat}{1}
	P_\text{AC} &= 4 \sqrt{2} G \varepsilon P_\text{cav} \frac{\sin \left(m_a \ell/2 \right)}{m_a}  \frac{t^C_\rightarrow  t^C_\uparrow}{1 - r_\uparrow^2} \,,
    \label{eqn:ACPower}
\end{alignat}
where we have assumed that all reflectivity coefficients are approximately 1. The sensitivity is estimated by finding the value of $g_{a\gamma\gamma}$ that sets the SNR to 1, with~\cite{Budker:2013hfa}
\begin{alignat}{1}
	\text{SNR} = \frac{P_\text{AC}}{S_\text{shot}^{1/2}} (T \tau)^{1/4} \,, 
    \label{eqn:SNR}
\end{alignat}
 where $S_\text{shot} = 2 P_\text{DC} \omega_0$ is the laser shot noise power spectral density with the DC power given by $P_\text{DC} = (2 \varepsilon t^C_\rightarrow)^2 P_\text{cav}$, $T$ is the integration time for this step in $\Delta \varphi$, $\tau \equiv 2\pi/(m_a v^2)$ is the coherence time of the axion field, and we assume $T \gg \tau$. The expression for the SNR in Eq.~(\ref{eqn:SNR}) is independent of $\varepsilon$; this is consistent with the fact that shot noise power always scales as the square root of the carrier power regardless of the amount of filtering performed, and is contrary to the result obtained in Ref.~\cite{Melissinos:2008vn}. 

Relative intensity noise (RIN) in our set-up is generated with the same polarization as the carrier and is orthogonal to our signal; it can be easily reduced by adjusting $\varepsilon$, balancing the photodetectors used to measure the signal, and using the transmitted carrier intensity to cancel this source of noise. RIN can also change the length of the cavity, but this can be suppressed with a high-bandwidth feedback loop.  

Another source of noise in our set-up is laser technical noise, which leads to finite laser frequency width and decreases the sensitivity of ADBC as $m_a \to 0$. In order to probe axion masses down to $m_a \sim 10^{-13}$ eV, technical noise must be subdominant to shot noise down to $\nu_a \sim $ 10--100 Hz. ADBC will adopt many of the same techniques used by the LIGO Collaboration to achieve isolation at these frequencies~\cite{Cook:2012tu}. A LIGO-like suspension system mounted on a rotating platform will be used for both pairs of mirrors. Since only a single beam is used in a cavity which is held on resonance via feedback to $\omega_0$, radiation pressure and other displacement noises are less relevant. Thermal noise in the mounted optics, for instance, will dominate over other non-technical noises (e.g.\ quantum radiation pressure noise),
with an estimated magnitude of~\cite{Gonzalez1994}
\begin{alignat}{1}
    S_{\Delta \varphi}^{1/2} \sim \frac{d \Delta \varphi}{d \theta} \frac{S_x^{1/2}}{\ell_{DB}} \sim 10^{-14} \frac{\SI{100}{Hz}}{\nu_a}\,.
    \label{eqn:thermal_noise}
\end{alignat}
This places requirements on the experimental design for small values of $m_a$ where $G/m_a \sim S_{\Delta \varphi}^{1/2}$ ($g_{a\gamma\gamma} \sim 10^{-14}\,\si{GeV^{-1}}$ at $m_a \sim 10^{-13}\,\si{eV}$).

Several steps can be taken to ascertain that a signal is indeed due to the presence of axion dark matter. First, the axion always produces a signal for both $\Delta \varphi = \pm m_a \ell$, even though the configuration of the mirrors may be very different in each case. Second, we can tune the cavity so that $\Delta \varphi_1 + \Delta \varphi_2 = 2 m_a \ell$, but with $\Delta \varphi_1 \neq \Delta \varphi_2$, where $\Delta \varphi_{1,2}$ are the phase differences generated at mirrors $A,C$ and $B,D$ respectively. With such a tuning, $P_\text{AC} \propto 1 + \cos(m_a \ell - \Delta \varphi_2)$. This dependence arises only due to the off-diagonal entries in the translation matrix $T$ shown in Eq.~(\ref{eqn:3x3_translation_matrix}), and serves as an explicit check that the signal is produced from the conversion of laser power from the carrier to the signal polarization.

The expected sensitivity for a \SI{2}{\meter}, $P_\text{cav} = \SI{10}{\kilo\watt}$ and a \SI{40}{m}, $P_\text{cav} = \SI{1}{\mega\watt}$ version of ADBC is given in Fig.~\ref{fig:ADBC_reach}. The \SI{2}{\meter} benchmark can currently be achieved in the laboratory, while the \SI{40}{\meter} version is similar in size to the \SI{40}{\meter} LIGO prototype at Caltech~\cite{Abramovici:1996dz} or the Fermilab Holometer~\cite{Chou:2015sle}, using an optical configuration similar to Advanced LIGO~\cite{TheLIGOScientific:2014jea}.  In order to cover the range $0 < m_a \lesssim \pi/(5\ell)$, the experiment must be run a number of times given by $\mathcal{F}_\uparrow/5 \sim 5 \times 10^{4}$, each with a different value of $\theta$.
$\mathcal{F}_\uparrow/5$ is chosen so that the peak of each resonance in $m_a$ falls on the half-maximum for the previous resonance, starting from $m_a = 10^{-13}$~eV. Given a total integration time of $T_\text{tot} = $ 30 days, we integrate each step for $T = \max(N_\ell \, \tau, \text{ 1 sec})$, where $N_\text{2\,m} = 35$ and $N_\text{40\,m} = 4$. This choice is equivalent to allocating the integration time logarithmically among bins of $m_a$, as recommended by Ref.~\cite{Chaudhuri:2018rqn}, and in agreement with Refs.~\cite{Chaudhuri:2014dla,Kahn:2016aff}. The envelope of the sensitivity to $g_{a\gamma\gamma}$ can be obtained analytically from Eq.~(\ref{eqn:SNR}), giving
\begin{alignat}{2}
	g_{a\gamma\gamma} &> && \,\, \SI{6.13e-11}{\per\giga\eV} \frac{N_\ell^{-1/4}}{\text{sinc}(m_a \ell/2)} \nonumber \\
    & && \times \left(\frac{\SI{0.3}{\giga\eV \per \centi\meter\cubed}}{\rho_\text{DM}} \frac{\SI{10}{\kilo\watt}}{P_\text{cav}} \frac{\SI{1.064}{\micro\meter}}{\lambda_0} \right)^{1/2} \nonumber \\
    & && \times \left( \frac{\SI{2}{\meter}}{\ell} \frac{10^{-3}}{t_\uparrow^C} \frac{10^5}{\mathcal{F}_\uparrow} \right) \sqrt{\frac{m_a}{\SI{e-13}{\eV}}} \,\,,
\end{alignat}
with $\lambda_0$ being the laser wavelength. For a given $m_a$, adding up the SNR in quadrature from every step may improve the reach by up to a factor of 2. A  \SI{40}{m} cavity with a circulating laser power of \SI{1}{MW} in the cavity improves upon CAST limits~\cite{Anastassopoulos:2017ftl} by almost four orders of magnitude for  $m_a \sim 10^{-13}$ eV. Ultimately, full-sized versions of ABRACADABRA~\cite{Kahn:2016aff,Ouellet:2018beu,Ouellet:2019tlz} and DM-Radio~\cite{Chaudhuri:2014dla,Battesti:2018bgc} may eventually cover much of the parameter space shown in Fig.~\ref{fig:ADBC_reach}. ADBC can, however, serve as a powerful complementary search to these experiments, relying on a completely different strategy in looking for axions. In particular, ADBC's ability to obtain two separate resonances at $\Delta \varphi = \pm m_a \ell$ is a striking experimental signature that would bolster any potential evidence for axions in other experiments.

\vspace{0.2cm}
\textit{Conclusion.}
We proposed a new axion interferometry experimental design that exploits the birefringence of a bowtie cavity in order to generate axion-modulated, vertically polarized sidebands from a horizontally polarized laser beam carrier. This design is practical to implement and can improve on the reach of previous interferometry designs from $m_a \sim 1/(\mathcal{F} \ell)$ up to $m_a \sim 1/\ell$, with the sensitivity improving with finesse. The sensitivity and mass range of our experiment can both be improved by a careful design of the mirrors used in the cavity, so that the cavity is optimally coupled with minimal loss, and the phase shift at each end extends to $\Delta \varphi = \pi/2$. We look forward to implementing this design and beginning the search for axions with the ADBC experiment.

\vspace{0.2cm}
\textit{Acknowledgments}. The authors would like to thank Anson Hook, Yonatan Kahn, Kaicheng Liang, and Benjamin Safdi for helpful discussions. The work of HL and JT was supported by the Office of High Energy Physics of the U.S. Department of Energy (DOE) under grant DE-SC-0012567.

\bibliography{axion_interferometry}

\begin{thebibliography}{52}%
\makeatletter
\providecommand \@ifxundefined [1]{%
 \@ifx{#1\undefined}
}%
\providecommand \@ifnum [1]{%
 \ifnum #1\expandafter \@firstoftwo
 \else \expandafter \@secondoftwo
 \fi
}%
\providecommand \@ifx [1]{%
 \ifx #1\expandafter \@firstoftwo
 \else \expandafter \@secondoftwo
 \fi
}%
\providecommand \natexlab [1]{#1}%
\providecommand \enquote  [1]{``#1''}%
\providecommand \bibnamefont  [1]{#1}%
\providecommand \bibfnamefont [1]{#1}%
\providecommand \citenamefont [1]{#1}%
\providecommand \href@noop [0]{\@secondoftwo}%
\providecommand \href [0]{\begingroup \@sanitize@url \@href}%
\providecommand \@href[1]{\@@startlink{#1}\@@href}%
\providecommand \@@href[1]{\endgroup#1\@@endlink}%
\providecommand \@sanitize@url [0]{\catcode `\\12\catcode `\$12\catcode
  `\&12\catcode `\#12\catcode `\^12\catcode `\_12\catcode `\%12\relax}%
\providecommand \@@startlink[1]{}%
\providecommand \@@endlink[0]{}%
\providecommand \url  [0]{\begingroup\@sanitize@url \@url }%
\providecommand \@url [1]{\endgroup\@href {#1}{\urlprefix }}%
\providecommand \urlprefix  [0]{URL }%
\providecommand \Eprint [0]{\href }%
\providecommand \doibase [0]{http://dx.doi.org/}%
\providecommand \selectlanguage [0]{\@gobble}%
\providecommand \bibinfo  [0]{\@secondoftwo}%
\providecommand \bibfield  [0]{\@secondoftwo}%
\providecommand \translation [1]{[#1]}%
\providecommand \BibitemOpen [0]{}%
\providecommand \bibitemStop [0]{}%
\providecommand \bibitemNoStop [0]{.\EOS\space}%
\providecommand \EOS [0]{\spacefactor3000\relax}%
\providecommand \BibitemShut  [1]{\csname bibitem#1\endcsname}%
\let\auto@bib@innerbib\@empty
\bibitem [{\citenamefont {Peccei}\ and\ \citenamefont
  {Quinn}(1977{\natexlab{a}})}]{Peccei:1977hh}%
  \BibitemOpen
  \bibfield  {author} {\bibinfo {author} {\bibfnamefont {R.~D.}\ \bibnamefont
  {Peccei}}\ and\ \bibinfo {author} {\bibfnamefont {Helen~R.}\ \bibnamefont
  {Quinn}},\ }\bibfield  {title} {\enquote {\bibinfo {title} {{CP Conservation
  in the Presence of Instantons}},}\ }\href {\doibase
  10.1103/PhysRevLett.38.1440} {\bibfield  {journal} {\bibinfo  {journal}
  {Phys. Rev. Lett.}\ }\textbf {\bibinfo {volume} {38}},\ \bibinfo {pages}
  {1440--1443} (\bibinfo {year} {1977}{\natexlab{a}})}\BibitemShut {NoStop}%
\bibitem [{\citenamefont {Peccei}\ and\ \citenamefont
  {Quinn}(1977{\natexlab{b}})}]{Peccei:1977ur}%
  \BibitemOpen
  \bibfield  {author} {\bibinfo {author} {\bibfnamefont {R.~D.}\ \bibnamefont
  {Peccei}}\ and\ \bibinfo {author} {\bibfnamefont {Helen~R.}\ \bibnamefont
  {Quinn}},\ }\bibfield  {title} {\enquote {\bibinfo {title} {{Constraints
  Imposed by CP Conservation in the Presence of Instantons}},}\ }\href
  {\doibase 10.1103/PhysRevD.16.1791} {\bibfield  {journal} {\bibinfo
  {journal} {Phys. Rev.}\ }\textbf {\bibinfo {volume} {D16}},\ \bibinfo {pages}
  {1791--1797} (\bibinfo {year} {1977}{\natexlab{b}})}\BibitemShut {NoStop}%
\bibitem [{\citenamefont {Weinberg}(1978)}]{Weinberg:1977ma}%
  \BibitemOpen
  \bibfield  {author} {\bibinfo {author} {\bibfnamefont {Steven}\ \bibnamefont
  {Weinberg}},\ }\bibfield  {title} {\enquote {\bibinfo {title} {{A New Light
  Boson?}}}\ }\href {\doibase 10.1103/PhysRevLett.40.223} {\bibfield  {journal}
  {\bibinfo  {journal} {Phys. Rev. Lett.}\ }\textbf {\bibinfo {volume} {40}},\
  \bibinfo {pages} {223--226} (\bibinfo {year} {1978})}\BibitemShut {NoStop}%
\bibitem [{\citenamefont {Wilczek}(1978)}]{Wilczek:1977pj}%
  \BibitemOpen
  \bibfield  {author} {\bibinfo {author} {\bibfnamefont {Frank}\ \bibnamefont
  {Wilczek}},\ }\bibfield  {title} {\enquote {\bibinfo {title} {{Problem of
  Strong p and t Invariance in the Presence of Instantons}},}\ }\href {\doibase
  10.1103/PhysRevLett.40.279} {\bibfield  {journal} {\bibinfo  {journal} {Phys.
  Rev. Lett.}\ }\textbf {\bibinfo {volume} {40}},\ \bibinfo {pages} {279--282}
  (\bibinfo {year} {1978})}\BibitemShut {NoStop}%
\bibitem [{\citenamefont {Preskill}\ \emph {et~al.}(1983)\citenamefont
  {Preskill}, \citenamefont {Wise},\ and\ \citenamefont
  {Wilczek}}]{Preskill:1982cy}%
  \BibitemOpen
  \bibfield  {author} {\bibinfo {author} {\bibfnamefont {John}\ \bibnamefont
  {Preskill}}, \bibinfo {author} {\bibfnamefont {Mark~B.}\ \bibnamefont
  {Wise}}, \ and\ \bibinfo {author} {\bibfnamefont {Frank}\ \bibnamefont
  {Wilczek}},\ }\bibfield  {title} {\enquote {\bibinfo {title} {{Cosmology of
  the Invisible Axion}},}\ }\href {\doibase 10.1016/0370-2693(83)90637-8}
  {\bibfield  {journal} {\bibinfo  {journal} {Phys. Lett.}\ }\textbf {\bibinfo
  {volume} {B120}},\ \bibinfo {pages} {127--132} (\bibinfo {year}
  {1983})}\BibitemShut {NoStop}%
\bibitem [{\citenamefont {Abbott}\ and\ \citenamefont
  {Sikivie}(1983)}]{Abbott:1982af}%
  \BibitemOpen
  \bibfield  {author} {\bibinfo {author} {\bibfnamefont {L.~F.}\ \bibnamefont
  {Abbott}}\ and\ \bibinfo {author} {\bibfnamefont {P.}~\bibnamefont
  {Sikivie}},\ }\bibfield  {title} {\enquote {\bibinfo {title} {{A Cosmological
  Bound on the Invisible Axion}},}\ }\href {\doibase
  10.1016/0370-2693(83)90638-X} {\bibfield  {journal} {\bibinfo  {journal}
  {Phys. Lett.}\ }\textbf {\bibinfo {volume} {B120}},\ \bibinfo {pages}
  {133--136} (\bibinfo {year} {1983})}\BibitemShut {NoStop}%
\bibitem [{\citenamefont {Dine}\ and\ \citenamefont
  {Fischler}(1983)}]{Dine:1982ah}%
  \BibitemOpen
  \bibfield  {author} {\bibinfo {author} {\bibfnamefont {Michael}\ \bibnamefont
  {Dine}}\ and\ \bibinfo {author} {\bibfnamefont {Willy}\ \bibnamefont
  {Fischler}},\ }\bibfield  {title} {\enquote {\bibinfo {title} {{The Not So
  Harmless Axion}},}\ }\href {\doibase 10.1016/0370-2693(83)90639-1} {\bibfield
   {journal} {\bibinfo  {journal} {Phys. Lett.}\ }\textbf {\bibinfo {volume}
  {B120}},\ \bibinfo {pages} {137--141} (\bibinfo {year} {1983})}\BibitemShut
  {NoStop}%
\bibitem [{\citenamefont {Graham}\ \emph {et~al.}(2015)\citenamefont {Graham},
  \citenamefont {Irastorza}, \citenamefont {Lamoreaux}, \citenamefont
  {Lindner},\ and\ \citenamefont {van Bibber}}]{Graham:2015ouw}%
  \BibitemOpen
  \bibfield  {author} {\bibinfo {author} {\bibfnamefont {Peter~W.}\
  \bibnamefont {Graham}}, \bibinfo {author} {\bibfnamefont {Igor~G.}\
  \bibnamefont {Irastorza}}, \bibinfo {author} {\bibfnamefont {Steven~K.}\
  \bibnamefont {Lamoreaux}}, \bibinfo {author} {\bibfnamefont {Axel}\
  \bibnamefont {Lindner}}, \ and\ \bibinfo {author} {\bibfnamefont {Karl~A.}\
  \bibnamefont {van Bibber}},\ }\bibfield  {title} {\enquote {\bibinfo {title}
  {{Experimental Searches for the Axion and Axion-Like Particles}},}\ }\href
  {\doibase 10.1146/annurev-nucl-102014-022120} {\bibfield  {journal} {\bibinfo
   {journal} {Ann. Rev. Nucl. Part. Sci.}\ }\textbf {\bibinfo {volume} {65}},\
  \bibinfo {pages} {485--514} (\bibinfo {year} {2015})},\ \Eprint
  {http://arxiv.org/abs/1602.00039} {arXiv:1602.00039 [hep-ex]} \BibitemShut
  {NoStop}%
\bibitem [{\citenamefont {Jaeckel}\ and\ \citenamefont
  {Ringwald}(2010)}]{Jaeckel:2010ni}%
  \BibitemOpen
  \bibfield  {author} {\bibinfo {author} {\bibfnamefont {Joerg}\ \bibnamefont
  {Jaeckel}}\ and\ \bibinfo {author} {\bibfnamefont {Andreas}\ \bibnamefont
  {Ringwald}},\ }\bibfield  {title} {\enquote {\bibinfo {title} {{The
  Low-Energy Frontier of Particle Physics}},}\ }\href {\doibase
  10.1146/annurev.nucl.012809.104433} {\bibfield  {journal} {\bibinfo
  {journal} {Ann. Rev. Nucl. Part. Sci.}\ }\textbf {\bibinfo {volume} {60}},\
  \bibinfo {pages} {405--437} (\bibinfo {year} {2010})},\ \Eprint
  {http://arxiv.org/abs/1002.0329} {arXiv:1002.0329 [hep-ph]} \BibitemShut
  {NoStop}%
\bibitem [{\citenamefont {Svrcek}\ and\ \citenamefont
  {Witten}(2006)}]{Svrcek:2006yi}%
  \BibitemOpen
  \bibfield  {author} {\bibinfo {author} {\bibfnamefont {Peter}\ \bibnamefont
  {Svrcek}}\ and\ \bibinfo {author} {\bibfnamefont {Edward}\ \bibnamefont
  {Witten}},\ }\bibfield  {title} {\enquote {\bibinfo {title} {{Axions In
  String Theory}},}\ }\href {\doibase 10.1088/1126-6708/2006/06/051} {\bibfield
   {journal} {\bibinfo  {journal} {JHEP}\ }\textbf {\bibinfo {volume} {06}},\
  \bibinfo {pages} {051} (\bibinfo {year} {2006})},\ \Eprint
  {http://arxiv.org/abs/hep-th/0605206} {arXiv:hep-th/0605206 [hep-th]}
  \BibitemShut {NoStop}%
\bibitem [{\citenamefont {Arvanitaki}\ \emph {et~al.}(2010)\citenamefont
  {Arvanitaki}, \citenamefont {Dimopoulos}, \citenamefont {Dubovsky},
  \citenamefont {Kaloper},\ and\ \citenamefont
  {March-Russell}}]{Arvanitaki:2009fg}%
  \BibitemOpen
  \bibfield  {author} {\bibinfo {author} {\bibfnamefont {Asimina}\ \bibnamefont
  {Arvanitaki}}, \bibinfo {author} {\bibfnamefont {Savas}\ \bibnamefont
  {Dimopoulos}}, \bibinfo {author} {\bibfnamefont {Sergei}\ \bibnamefont
  {Dubovsky}}, \bibinfo {author} {\bibfnamefont {Nemanja}\ \bibnamefont
  {Kaloper}}, \ and\ \bibinfo {author} {\bibfnamefont {John}\ \bibnamefont
  {March-Russell}},\ }\bibfield  {title} {\enquote {\bibinfo {title} {{String
  Axiverse}},}\ }\href {\doibase 10.1103/PhysRevD.81.123530} {\bibfield
  {journal} {\bibinfo  {journal} {Phys. Rev.}\ }\textbf {\bibinfo {volume}
  {D81}},\ \bibinfo {pages} {123530} (\bibinfo {year} {2010})},\ \Eprint
  {http://arxiv.org/abs/0905.4720} {arXiv:0905.4720 [hep-th]} \BibitemShut
  {NoStop}%
\bibitem [{\citenamefont {Acharya}\ \emph {et~al.}(2010)\citenamefont
  {Acharya}, \citenamefont {Bobkov},\ and\ \citenamefont
  {Kumar}}]{Acharya:2010zx}%
  \BibitemOpen
  \bibfield  {author} {\bibinfo {author} {\bibfnamefont {Bobby~Samir}\
  \bibnamefont {Acharya}}, \bibinfo {author} {\bibfnamefont {Konstantin}\
  \bibnamefont {Bobkov}}, \ and\ \bibinfo {author} {\bibfnamefont {Piyush}\
  \bibnamefont {Kumar}},\ }\bibfield  {title} {\enquote {\bibinfo {title} {{An
  M Theory Solution to the Strong CP Problem and Constraints on the
  Axiverse}},}\ }\href {\doibase 10.1007/JHEP11(2010)105} {\bibfield  {journal}
  {\bibinfo  {journal} {JHEP}\ }\textbf {\bibinfo {volume} {11}},\ \bibinfo
  {pages} {105} (\bibinfo {year} {2010})},\ \Eprint
  {http://arxiv.org/abs/1004.5138} {arXiv:1004.5138 [hep-th]} \BibitemShut
  {NoStop}%
\bibitem [{\citenamefont {Cicoli}\ \emph {et~al.}(2012)\citenamefont {Cicoli},
  \citenamefont {Goodsell},\ and\ \citenamefont {Ringwald}}]{Cicoli:2012sz}%
  \BibitemOpen
  \bibfield  {author} {\bibinfo {author} {\bibfnamefont {Michele}\ \bibnamefont
  {Cicoli}}, \bibinfo {author} {\bibfnamefont {Mark}\ \bibnamefont {Goodsell}},
  \ and\ \bibinfo {author} {\bibfnamefont {Andreas}\ \bibnamefont {Ringwald}},\
  }\bibfield  {title} {\enquote {\bibinfo {title} {{The type IIB string
  axiverse and its low-energy phenomenology}},}\ }\href {\doibase
  10.1007/JHEP10(2012)146} {\bibfield  {journal} {\bibinfo  {journal} {JHEP}\
  }\textbf {\bibinfo {volume} {10}},\ \bibinfo {pages} {146} (\bibinfo {year}
  {2012})},\ \Eprint {http://arxiv.org/abs/1206.0819} {arXiv:1206.0819
  [hep-th]} \BibitemShut {NoStop}%
\bibitem [{\citenamefont {Sikivie}(1983)}]{Sikivie:1983ip}%
  \BibitemOpen
  \bibfield  {author} {\bibinfo {author} {\bibfnamefont {P.}~\bibnamefont
  {Sikivie}},\ }\bibfield  {title} {\enquote {\bibinfo {title} {{Experimental
  Tests of the Invisible Axion}},}\ }\bibfield  {booktitle} {\emph {\bibinfo
  {booktitle} {{Particle physics and cosmology: Dark matter}}},\ }\href
  {\doibase 10.1103/PhysRevLett.51.1415, 10.1103/PhysRevLett.52.695.2}
  {\bibfield  {journal} {\bibinfo  {journal} {Phys. Rev. Lett.}\ }\textbf
  {\bibinfo {volume} {51}},\ \bibinfo {pages} {1415--1417} (\bibinfo {year}
  {1983})}\BibitemShut {NoStop}%
\bibitem [{\citenamefont {Wilczek}(1987)}]{Wilczek:1987mv}%
  \BibitemOpen
  \bibfield  {author} {\bibinfo {author} {\bibfnamefont {Frank}\ \bibnamefont
  {Wilczek}},\ }\bibfield  {title} {\enquote {\bibinfo {title} {{Two
  Applications of Axion Electrodynamics}},}\ }\href {\doibase
  10.1103/PhysRevLett.58.1799} {\bibfield  {journal} {\bibinfo  {journal}
  {Phys. Rev. Lett.}\ }\textbf {\bibinfo {volume} {58}},\ \bibinfo {pages}
  {1799} (\bibinfo {year} {1987})}\BibitemShut {NoStop}%
\bibitem [{\citenamefont {Cameron}\ \emph {et~al.}(1993)\citenamefont {Cameron}
  \emph {et~al.}}]{Cameron:1993mr}%
  \BibitemOpen
  \bibfield  {author} {\bibinfo {author} {\bibfnamefont {R.}~\bibnamefont
  {Cameron}} \emph {et~al.},\ }\bibfield  {title} {\enquote {\bibinfo {title}
  {{Search for nearly massless, weakly coupled particles by optical
  techniques}},}\ }\href {\doibase 10.1103/PhysRevD.47.3707} {\bibfield
  {journal} {\bibinfo  {journal} {Phys. Rev.}\ }\textbf {\bibinfo {volume}
  {D47}},\ \bibinfo {pages} {3707--3725} (\bibinfo {year} {1993})}\BibitemShut
  {NoStop}%
\bibitem [{\citenamefont {Tam}\ and\ \citenamefont {Yang}(2012)}]{Tam:2011kw}%
  \BibitemOpen
  \bibfield  {author} {\bibinfo {author} {\bibfnamefont {H.}~\bibnamefont
  {Tam}}\ and\ \bibinfo {author} {\bibfnamefont {Q.}~\bibnamefont {Yang}},\
  }\bibfield  {title} {\enquote {\bibinfo {title} {{Production and Detection of
  Axion-like Particles by Interferometry}},}\ }\href {\doibase
  10.1016/j.physletb.2012.08.050} {\bibfield  {journal} {\bibinfo  {journal}
  {Phys. Lett.}\ }\textbf {\bibinfo {volume} {B716}},\ \bibinfo {pages}
  {435--440} (\bibinfo {year} {2012})},\ \Eprint
  {http://arxiv.org/abs/1107.1712} {arXiv:1107.1712 [hep-ph]} \BibitemShut
  {NoStop}%
\bibitem [{\citenamefont {Della~Valle}\ \emph {et~al.}(2016)\citenamefont
  {Della~Valle}, \citenamefont {Ejlli}, \citenamefont {Gastaldi}, \citenamefont
  {Messineo}, \citenamefont {Milotti}, \citenamefont {Pengo}, \citenamefont
  {Ruoso},\ and\ \citenamefont {Zavattini}}]{DellaValle:2015xxa}%
  \BibitemOpen
  \bibfield  {author} {\bibinfo {author} {\bibfnamefont {Federico}\
  \bibnamefont {Della~Valle}}, \bibinfo {author} {\bibfnamefont {Aldo}\
  \bibnamefont {Ejlli}}, \bibinfo {author} {\bibfnamefont {Ugo}\ \bibnamefont
  {Gastaldi}}, \bibinfo {author} {\bibfnamefont {Giuseppe}\ \bibnamefont
  {Messineo}}, \bibinfo {author} {\bibfnamefont {Edoardo}\ \bibnamefont
  {Milotti}}, \bibinfo {author} {\bibfnamefont {Ruggero}\ \bibnamefont
  {Pengo}}, \bibinfo {author} {\bibfnamefont {Giuseppe}\ \bibnamefont {Ruoso}},
  \ and\ \bibinfo {author} {\bibfnamefont {Guido}\ \bibnamefont {Zavattini}},\
  }\bibfield  {title} {\enquote {\bibinfo {title} {{The PVLAS experiment:
  measuring vacuum magnetic birefringence and dichroism with a birefringent
  Fabry–Perot cavity}},}\ }\href {\doibase 10.1140/epjc/s10052-015-3869-8}
  {\bibfield  {journal} {\bibinfo  {journal} {Eur. Phys. J.}\ }\textbf
  {\bibinfo {volume} {C76}},\ \bibinfo {pages} {24} (\bibinfo {year} {2016})},\
  \Eprint {http://arxiv.org/abs/1510.08052} {arXiv:1510.08052 [physics.optics]}
  \BibitemShut {NoStop}%
\bibitem [{\citenamefont {Chou}\ \emph {et~al.}(2008)\citenamefont {Chou},
  \citenamefont {Wester}, \citenamefont {Baumbaugh}, \citenamefont {Gustafson},
  \citenamefont {Irizarry-Valle}, \citenamefont {Mazur}, \citenamefont
  {Steffen}, \citenamefont {Tomlin}, \citenamefont {Yang},\ and\ \citenamefont
  {Yoo}}]{Chou:2007zzc}%
  \BibitemOpen
  \bibfield  {author} {\bibinfo {author} {\bibfnamefont {Aaron~S.}\
  \bibnamefont {Chou}}, \bibinfo {author} {\bibfnamefont {William~Carl}\
  \bibnamefont {Wester}, \bibfnamefont {III}}, \bibinfo {author} {\bibfnamefont
  {A.}~\bibnamefont {Baumbaugh}}, \bibinfo {author} {\bibfnamefont
  {H.~Richard}\ \bibnamefont {Gustafson}}, \bibinfo {author} {\bibfnamefont
  {Y.}~\bibnamefont {Irizarry-Valle}}, \bibinfo {author} {\bibfnamefont
  {P.~O.}\ \bibnamefont {Mazur}}, \bibinfo {author} {\bibfnamefont {Jason~H.}\
  \bibnamefont {Steffen}}, \bibinfo {author} {\bibfnamefont {R.}~\bibnamefont
  {Tomlin}}, \bibinfo {author} {\bibfnamefont {X.}~\bibnamefont {Yang}}, \ and\
  \bibinfo {author} {\bibfnamefont {J.}~\bibnamefont {Yoo}} (\bibinfo
  {collaboration} {GammeV (T-969)}),\ }\bibfield  {title} {\enquote {\bibinfo
  {title} {{Search for axion-like particles using a variable baseline photon
  regeneration technique}},}\ }\href {\doibase 10.1103/PhysRevLett.100.080402}
  {\bibfield  {journal} {\bibinfo  {journal} {Phys. Rev. Lett.}\ }\textbf
  {\bibinfo {volume} {100}},\ \bibinfo {pages} {080402} (\bibinfo {year}
  {2008})},\ \Eprint {http://arxiv.org/abs/0710.3783} {arXiv:0710.3783
  [hep-ex]} \BibitemShut {NoStop}%
\bibitem [{\citenamefont {Robilliard}\ \emph {et~al.}(2007)\citenamefont
  {Robilliard}, \citenamefont {Battesti}, \citenamefont {Fouche}, \citenamefont
  {Mauchain}, \citenamefont {Sautivet}, \citenamefont {Amiranoff},\ and\
  \citenamefont {Rizzo}}]{Robilliard:2007bq}%
  \BibitemOpen
  \bibfield  {author} {\bibinfo {author} {\bibfnamefont {Cecile}\ \bibnamefont
  {Robilliard}}, \bibinfo {author} {\bibfnamefont {Remy}\ \bibnamefont
  {Battesti}}, \bibinfo {author} {\bibfnamefont {Mathilde}\ \bibnamefont
  {Fouche}}, \bibinfo {author} {\bibfnamefont {Julien}\ \bibnamefont
  {Mauchain}}, \bibinfo {author} {\bibfnamefont {Anne-Marie}\ \bibnamefont
  {Sautivet}}, \bibinfo {author} {\bibfnamefont {Francois}\ \bibnamefont
  {Amiranoff}}, \ and\ \bibinfo {author} {\bibfnamefont {Carlo}\ \bibnamefont
  {Rizzo}},\ }\bibfield  {title} {\enquote {\bibinfo {title} {{No light shining
  through a wall}},}\ }\href {\doibase 10.1103/PhysRevLett.99.190403}
  {\bibfield  {journal} {\bibinfo  {journal} {Phys. Rev. Lett.}\ }\textbf
  {\bibinfo {volume} {99}},\ \bibinfo {pages} {190403} (\bibinfo {year}
  {2007})},\ \Eprint {http://arxiv.org/abs/0707.1296} {arXiv:0707.1296
  [hep-ex]} \BibitemShut {NoStop}%
\bibitem [{\citenamefont {Ehret}\ \emph {et~al.}(2010)\citenamefont {Ehret}
  \emph {et~al.}}]{Ehret:2010mh}%
  \BibitemOpen
  \bibfield  {author} {\bibinfo {author} {\bibfnamefont {Klaus}\ \bibnamefont
  {Ehret}} \emph {et~al.},\ }\bibfield  {title} {\enquote {\bibinfo {title}
  {{New ALPS Results on Hidden-Sector Lightweights}},}\ }\href {\doibase
  10.1016/j.physletb.2010.04.066} {\bibfield  {journal} {\bibinfo  {journal}
  {Phys. Lett.}\ }\textbf {\bibinfo {volume} {B689}},\ \bibinfo {pages}
  {149--155} (\bibinfo {year} {2010})},\ \Eprint
  {http://arxiv.org/abs/1004.1313} {arXiv:1004.1313 [hep-ex]} \BibitemShut
  {NoStop}%
\bibitem [{\citenamefont {Betz}\ \emph {et~al.}(2013)\citenamefont {Betz},
  \citenamefont {Caspers}, \citenamefont {Gasior}, \citenamefont {Thumm},\ and\
  \citenamefont {Rieger}}]{Betz:2013dza}%
  \BibitemOpen
  \bibfield  {author} {\bibinfo {author} {\bibfnamefont {M.}~\bibnamefont
  {Betz}}, \bibinfo {author} {\bibfnamefont {F.}~\bibnamefont {Caspers}},
  \bibinfo {author} {\bibfnamefont {M.}~\bibnamefont {Gasior}}, \bibinfo
  {author} {\bibfnamefont {M.}~\bibnamefont {Thumm}}, \ and\ \bibinfo {author}
  {\bibfnamefont {S.~W.}\ \bibnamefont {Rieger}},\ }\bibfield  {title}
  {\enquote {\bibinfo {title} {{First results of the CERN Resonant Weakly
  Interacting sub-eV Particle Search (CROWS)}},}\ }\href {\doibase
  10.1103/PhysRevD.88.075014} {\bibfield  {journal} {\bibinfo  {journal} {Phys.
  Rev.}\ }\textbf {\bibinfo {volume} {D88}},\ \bibinfo {pages} {075014}
  (\bibinfo {year} {2013})},\ \Eprint {http://arxiv.org/abs/1310.8098}
  {arXiv:1310.8098 [physics.ins-det]} \BibitemShut {NoStop}%
\bibitem [{\citenamefont {Ballou}\ \emph {et~al.}(2014)\citenamefont {Ballou}
  \emph {et~al.}}]{Ballou:2014myz}%
  \BibitemOpen
  \bibfield  {author} {\bibinfo {author} {\bibfnamefont {Rafik}\ \bibnamefont
  {Ballou}} \emph {et~al.},\ }\bibfield  {title} {\enquote {\bibinfo {title}
  {{Latest Results of the OSQAR Photon Regeneration Experiment for Axion-Like
  Particle Search}},}\ }in\ \href {\doibase
  10.3204/DESY-PROC-2014-03/pugnat_pierre} {\emph {\bibinfo {booktitle}
  {{Proceedings, 10th Patras Workshop on Axions, WIMPs and WISPs (AXION-WIMP
  2014): Geneva, Switzerland, June 29-July 4, 2014}}}}\ (\bibinfo {year}
  {2014})\ pp.\ \bibinfo {pages} {125--130},\ \Eprint
  {http://arxiv.org/abs/1410.2566} {arXiv:1410.2566 [hep-ex]} \BibitemShut
  {NoStop}%
\bibitem [{\citenamefont {Ballou}\ \emph {et~al.}(2015)\citenamefont {Ballou}
  \emph {et~al.}}]{Ballou:2015cka}%
  \BibitemOpen
  \bibfield  {author} {\bibinfo {author} {\bibfnamefont {R.}~\bibnamefont
  {Ballou}} \emph {et~al.} (\bibinfo {collaboration} {OSQAR}),\ }\bibfield
  {title} {\enquote {\bibinfo {title} {{New exclusion limits on scalar and
  pseudoscalar axionlike particles from light shining through a wall}},}\
  }\href {\doibase 10.1103/PhysRevD.92.092002} {\bibfield  {journal} {\bibinfo
  {journal} {Phys. Rev.}\ }\textbf {\bibinfo {volume} {D92}},\ \bibinfo {pages}
  {092002} (\bibinfo {year} {2015})},\ \Eprint
  {http://arxiv.org/abs/1506.08082} {arXiv:1506.08082 [hep-ex]} \BibitemShut
  {NoStop}%
\bibitem [{\citenamefont {Horns}\ \emph {et~al.}(2013)\citenamefont {Horns},
  \citenamefont {Jaeckel}, \citenamefont {Lindner}, \citenamefont {Lobanov},
  \citenamefont {Redondo},\ and\ \citenamefont {Ringwald}}]{Horns:2012jf}%
  \BibitemOpen
  \bibfield  {author} {\bibinfo {author} {\bibfnamefont {Dieter}\ \bibnamefont
  {Horns}}, \bibinfo {author} {\bibfnamefont {Joerg}\ \bibnamefont {Jaeckel}},
  \bibinfo {author} {\bibfnamefont {Axel}\ \bibnamefont {Lindner}}, \bibinfo
  {author} {\bibfnamefont {Andrei}\ \bibnamefont {Lobanov}}, \bibinfo {author}
  {\bibfnamefont {Javier}\ \bibnamefont {Redondo}}, \ and\ \bibinfo {author}
  {\bibfnamefont {Andreas}\ \bibnamefont {Ringwald}},\ }\bibfield  {title}
  {\enquote {\bibinfo {title} {{Searching for WISPy Cold Dark Matter with a
  Dish Antenna}},}\ }\href {\doibase 10.1088/1475-7516/2013/04/016} {\bibfield
  {journal} {\bibinfo  {journal} {JCAP}\ }\textbf {\bibinfo {volume} {1304}},\
  \bibinfo {pages} {016} (\bibinfo {year} {2013})},\ \Eprint
  {http://arxiv.org/abs/1212.2970} {arXiv:1212.2970 [hep-ph]} \BibitemShut
  {NoStop}%
\bibitem [{\citenamefont {Chaudhuri}\ \emph {et~al.}(2015)\citenamefont
  {Chaudhuri}, \citenamefont {Graham}, \citenamefont {Irwin}, \citenamefont
  {Mardon}, \citenamefont {Rajendran},\ and\ \citenamefont
  {Zhao}}]{Chaudhuri:2014dla}%
  \BibitemOpen
  \bibfield  {author} {\bibinfo {author} {\bibfnamefont {Saptarshi}\
  \bibnamefont {Chaudhuri}}, \bibinfo {author} {\bibfnamefont {Peter~W.}\
  \bibnamefont {Graham}}, \bibinfo {author} {\bibfnamefont {Kent}\ \bibnamefont
  {Irwin}}, \bibinfo {author} {\bibfnamefont {Jeremy}\ \bibnamefont {Mardon}},
  \bibinfo {author} {\bibfnamefont {Surjeet}\ \bibnamefont {Rajendran}}, \ and\
  \bibinfo {author} {\bibfnamefont {Yue}\ \bibnamefont {Zhao}},\ }\bibfield
  {title} {\enquote {\bibinfo {title} {{Radio for hidden-photon dark matter
  detection}},}\ }\href {\doibase 10.1103/PhysRevD.92.075012} {\bibfield
  {journal} {\bibinfo  {journal} {Phys. Rev.}\ }\textbf {\bibinfo {volume}
  {D92}},\ \bibinfo {pages} {075012} (\bibinfo {year} {2015})},\ \Eprint
  {http://arxiv.org/abs/1411.7382} {arXiv:1411.7382 [hep-ph]} \BibitemShut
  {NoStop}%
\bibitem [{\citenamefont {Kahn}\ \emph {et~al.}(2016)\citenamefont {Kahn},
  \citenamefont {Safdi},\ and\ \citenamefont {Thaler}}]{Kahn:2016aff}%
  \BibitemOpen
  \bibfield  {author} {\bibinfo {author} {\bibfnamefont {Yonatan}\ \bibnamefont
  {Kahn}}, \bibinfo {author} {\bibfnamefont {Benjamin~R.}\ \bibnamefont
  {Safdi}}, \ and\ \bibinfo {author} {\bibfnamefont {Jesse}\ \bibnamefont
  {Thaler}},\ }\bibfield  {title} {\enquote {\bibinfo {title} {{Broadband and
  Resonant Approaches to Axion Dark Matter Detection}},}\ }\href {\doibase
  10.1103/PhysRevLett.117.141801} {\bibfield  {journal} {\bibinfo  {journal}
  {Phys. Rev. Lett.}\ }\textbf {\bibinfo {volume} {117}},\ \bibinfo {pages}
  {141801} (\bibinfo {year} {2016})},\ \Eprint
  {http://arxiv.org/abs/1602.01086} {arXiv:1602.01086 [hep-ph]} \BibitemShut
  {NoStop}%
\bibitem [{\citenamefont {Caldwell}\ \emph {et~al.}(2017)\citenamefont
  {Caldwell}, \citenamefont {Dvali}, \citenamefont {Majorovits}, \citenamefont
  {Millar}, \citenamefont {Raffelt}, \citenamefont {Redondo}, \citenamefont
  {Reimann}, \citenamefont {Simon},\ and\ \citenamefont
  {Steffen}}]{TheMADMAXWorkingGroup:2016hpc}%
  \BibitemOpen
  \bibfield  {author} {\bibinfo {author} {\bibfnamefont {Allen}\ \bibnamefont
  {Caldwell}}, \bibinfo {author} {\bibfnamefont {Gia}\ \bibnamefont {Dvali}},
  \bibinfo {author} {\bibfnamefont {Béla}\ \bibnamefont {Majorovits}},
  \bibinfo {author} {\bibfnamefont {Alexander}\ \bibnamefont {Millar}},
  \bibinfo {author} {\bibfnamefont {Georg}\ \bibnamefont {Raffelt}}, \bibinfo
  {author} {\bibfnamefont {Javier}\ \bibnamefont {Redondo}}, \bibinfo {author}
  {\bibfnamefont {Olaf}\ \bibnamefont {Reimann}}, \bibinfo {author}
  {\bibfnamefont {Frank}\ \bibnamefont {Simon}}, \ and\ \bibinfo {author}
  {\bibfnamefont {Frank}\ \bibnamefont {Steffen}} (\bibinfo {collaboration}
  {MADMAX Working Group}),\ }\bibfield  {title} {\enquote {\bibinfo {title}
  {{Dielectric Haloscopes: A New Way to Detect Axion Dark Matter}},}\ }\href
  {\doibase 10.1103/PhysRevLett.118.091801} {\bibfield  {journal} {\bibinfo
  {journal} {Phys. Rev. Lett.}\ }\textbf {\bibinfo {volume} {118}},\ \bibinfo
  {pages} {091801} (\bibinfo {year} {2017})},\ \Eprint
  {http://arxiv.org/abs/1611.05865} {arXiv:1611.05865 [physics.ins-det]}
  \BibitemShut {NoStop}%
\bibitem [{\citenamefont {Foster}\ \emph {et~al.}(2018)\citenamefont {Foster},
  \citenamefont {Rodd},\ and\ \citenamefont {Safdi}}]{Foster:2017hbq}%
  \BibitemOpen
  \bibfield  {author} {\bibinfo {author} {\bibfnamefont {Joshua~W.}\
  \bibnamefont {Foster}}, \bibinfo {author} {\bibfnamefont {Nicholas~L.}\
  \bibnamefont {Rodd}}, \ and\ \bibinfo {author} {\bibfnamefont {Benjamin~R.}\
  \bibnamefont {Safdi}},\ }\bibfield  {title} {\enquote {\bibinfo {title}
  {{Revealing the Dark Matter Halo with Axion Direct Detection}},}\ }\href
  {\doibase 10.1103/PhysRevD.97.123006} {\bibfield  {journal} {\bibinfo
  {journal} {Phys. Rev.}\ }\textbf {\bibinfo {volume} {D97}},\ \bibinfo {pages}
  {123006} (\bibinfo {year} {2018})},\ \Eprint
  {http://arxiv.org/abs/1711.10489} {arXiv:1711.10489 [astro-ph.CO]}
  \BibitemShut {NoStop}%
\bibitem [{\citenamefont {Chaudhuri}\ \emph {et~al.}(2018)\citenamefont
  {Chaudhuri}, \citenamefont {Irwin}, \citenamefont {Graham},\ and\
  \citenamefont {Mardon}}]{Chaudhuri:2018rqn}%
  \BibitemOpen
  \bibfield  {author} {\bibinfo {author} {\bibfnamefont {Saptarshi}\
  \bibnamefont {Chaudhuri}}, \bibinfo {author} {\bibfnamefont {Kent}\
  \bibnamefont {Irwin}}, \bibinfo {author} {\bibfnamefont {Peter~W.}\
  \bibnamefont {Graham}}, \ and\ \bibinfo {author} {\bibfnamefont {Jeremy}\
  \bibnamefont {Mardon}},\ }\bibfield  {title} {\enquote {\bibinfo {title}
  {{Fundamental Limits of Electromagnetic Axion and Hidden-Photon Dark Matter
  Searches: Part I - The Quantum Limit}},}\ }\href@noop {} {\  (\bibinfo {year}
  {2018})},\ \Eprint {http://arxiv.org/abs/1803.01627} {arXiv:1803.01627
  [hep-ph]} \BibitemShut {NoStop}%
\bibitem [{\citenamefont {Du}\ \emph {et~al.}(2018)\citenamefont {Du} \emph
  {et~al.}}]{Du:2018uak}%
  \BibitemOpen
  \bibfield  {author} {\bibinfo {author} {\bibfnamefont {N.}~\bibnamefont {Du}}
  \emph {et~al.} (\bibinfo {collaboration} {ADMX}),\ }\bibfield  {title}
  {\enquote {\bibinfo {title} {{A Search for Invisible Axion Dark Matter with
  the Axion Dark Matter Experiment}},}\ }\href {\doibase
  10.1103/PhysRevLett.120.151301} {\bibfield  {journal} {\bibinfo  {journal}
  {Phys. Rev. Lett.}\ }\textbf {\bibinfo {volume} {120}},\ \bibinfo {pages}
  {151301} (\bibinfo {year} {2018})},\ \Eprint
  {http://arxiv.org/abs/1804.05750} {arXiv:1804.05750 [hep-ex]} \BibitemShut
  {NoStop}%
\bibitem [{\citenamefont {Baryakhtar}\ \emph {et~al.}(2018)\citenamefont
  {Baryakhtar}, \citenamefont {Huang},\ and\ \citenamefont
  {Lasenby}}]{Baryakhtar:2018doz}%
  \BibitemOpen
  \bibfield  {author} {\bibinfo {author} {\bibfnamefont {Masha}\ \bibnamefont
  {Baryakhtar}}, \bibinfo {author} {\bibfnamefont {Junwu}\ \bibnamefont
  {Huang}}, \ and\ \bibinfo {author} {\bibfnamefont {Robert}\ \bibnamefont
  {Lasenby}},\ }\bibfield  {title} {\enquote {\bibinfo {title} {{Axion and
  hidden photon dark matter detection with multilayer optical haloscopes}},}\
  }\href {\doibase 10.1103/PhysRevD.98.035006} {\bibfield  {journal} {\bibinfo
  {journal} {Phys. Rev.}\ }\textbf {\bibinfo {volume} {D98}},\ \bibinfo {pages}
  {035006} (\bibinfo {year} {2018})},\ \Eprint
  {http://arxiv.org/abs/1803.11455} {arXiv:1803.11455 [hep-ph]} \BibitemShut
  {NoStop}%
\bibitem [{\citenamefont {Ouellet}\ \emph
  {et~al.}(2019{\natexlab{a}})\citenamefont {Ouellet} \emph
  {et~al.}}]{Ouellet:2018beu}%
  \BibitemOpen
  \bibfield  {author} {\bibinfo {author} {\bibfnamefont {Jonathan~L.}\
  \bibnamefont {Ouellet}} \emph {et~al.},\ }\bibfield  {title} {\enquote
  {\bibinfo {title} {{First Results from ABRACADABRA-10 cm: A Search for
  Sub-$\mu$eV Axion Dark Matter}},}\ }\href {\doibase
  10.1103/PhysRevLett.122.121802} {\bibfield  {journal} {\bibinfo  {journal}
  {Phys. Rev. Lett.}\ }\textbf {\bibinfo {volume} {122}},\ \bibinfo {pages}
  {121802} (\bibinfo {year} {2019}{\natexlab{a}})},\ \Eprint
  {http://arxiv.org/abs/1810.12257} {arXiv:1810.12257 [hep-ex]} \BibitemShut
  {NoStop}%
\bibitem [{\citenamefont {Ouellet}\ \emph
  {et~al.}(2019{\natexlab{b}})\citenamefont {Ouellet} \emph
  {et~al.}}]{Ouellet:2019tlz}%
  \BibitemOpen
  \bibfield  {author} {\bibinfo {author} {\bibfnamefont {Jonathan~L.}\
  \bibnamefont {Ouellet}} \emph {et~al.},\ }\bibfield  {title} {\enquote
  {\bibinfo {title} {{Design and implementation of the ABRACADABRA-10 cm axion
  dark matter search}},}\ }\href {\doibase 10.1103/PhysRevD.99.052012}
  {\bibfield  {journal} {\bibinfo  {journal} {Phys. Rev.}\ }\textbf {\bibinfo
  {volume} {D99}},\ \bibinfo {pages} {052012} (\bibinfo {year}
  {2019}{\natexlab{b}})},\ \Eprint {http://arxiv.org/abs/1901.10652}
  {arXiv:1901.10652 [physics.ins-det]} \BibitemShut {NoStop}%
\bibitem [{\citenamefont {Anastassopoulos}\ \emph {et~al.}(2017)\citenamefont
  {Anastassopoulos} \emph {et~al.}}]{Anastassopoulos:2017ftl}%
  \BibitemOpen
  \bibfield  {author} {\bibinfo {author} {\bibfnamefont {V.}~\bibnamefont
  {Anastassopoulos}} \emph {et~al.} (\bibinfo {collaboration} {CAST}),\
  }\bibfield  {title} {\enquote {\bibinfo {title} {{New CAST Limit on the
  Axion-Photon Interaction}},}\ }\href {\doibase 10.1038/nphys4109} {\bibfield
  {journal} {\bibinfo  {journal} {Nature Phys.}\ }\textbf {\bibinfo {volume}
  {13}},\ \bibinfo {pages} {584--590} (\bibinfo {year} {2017})},\ \Eprint
  {http://arxiv.org/abs/1705.02290} {arXiv:1705.02290 [hep-ex]} \BibitemShut
  {NoStop}%
\bibitem [{\citenamefont {Armengaud}\ \emph {et~al.}(2014)\citenamefont
  {Armengaud} \emph {et~al.}}]{Armengaud:2014gea}%
  \BibitemOpen
  \bibfield  {author} {\bibinfo {author} {\bibfnamefont {E.}~\bibnamefont
  {Armengaud}} \emph {et~al.},\ }\bibfield  {title} {\enquote {\bibinfo {title}
  {{Conceptual Design of the International Axion Observatory (IAXO)}},}\ }\href
  {\doibase 10.1088/1748-0221/9/05/T05002} {\bibfield  {journal} {\bibinfo
  {journal} {JINST}\ }\textbf {\bibinfo {volume} {9}},\ \bibinfo {pages}
  {T05002} (\bibinfo {year} {2014})},\ \Eprint {http://arxiv.org/abs/1401.3233}
  {arXiv:1401.3233 [physics.ins-det]} \BibitemShut {NoStop}%
\bibitem [{\citenamefont {Melissinos}(2009)}]{Melissinos:2008vn}%
  \BibitemOpen
  \bibfield  {author} {\bibinfo {author} {\bibfnamefont {Adrian~C.}\
  \bibnamefont {Melissinos}},\ }\bibfield  {title} {\enquote {\bibinfo {title}
  {{Search for Cosmic Axions using an Optical Interferometer}},}\ }\href
  {\doibase 10.1103/PhysRevLett.102.202001} {\bibfield  {journal} {\bibinfo
  {journal} {Phys. Rev. Lett.}\ }\textbf {\bibinfo {volume} {102}},\ \bibinfo
  {pages} {202001} (\bibinfo {year} {2009})},\ \Eprint
  {http://arxiv.org/abs/0807.1092} {arXiv:0807.1092 [hep-ph]} \BibitemShut
  {NoStop}%
\bibitem [{\citenamefont {DeRocco}\ and\ \citenamefont
  {Hook}(2018)}]{DeRocco:2018jwe}%
  \BibitemOpen
  \bibfield  {author} {\bibinfo {author} {\bibfnamefont {William}\ \bibnamefont
  {DeRocco}}\ and\ \bibinfo {author} {\bibfnamefont {Anson}\ \bibnamefont
  {Hook}},\ }\bibfield  {title} {\enquote {\bibinfo {title} {{Axion
  interferometry}},}\ }\href {\doibase 10.1103/PhysRevD.98.035021} {\bibfield
  {journal} {\bibinfo  {journal} {Phys. Rev.}\ }\textbf {\bibinfo {volume}
  {D98}},\ \bibinfo {pages} {035021} (\bibinfo {year} {2018})},\ \Eprint
  {http://arxiv.org/abs/1802.07273} {arXiv:1802.07273 [hep-ph]} \BibitemShut
  {NoStop}%
\bibitem [{\citenamefont {Obata}\ \emph {et~al.}(2018)\citenamefont {Obata},
  \citenamefont {Fujita},\ and\ \citenamefont {Michimura}}]{Obata:2018vvr}%
  \BibitemOpen
  \bibfield  {author} {\bibinfo {author} {\bibfnamefont {Ippei}\ \bibnamefont
  {Obata}}, \bibinfo {author} {\bibfnamefont {Tomohiro}\ \bibnamefont
  {Fujita}}, \ and\ \bibinfo {author} {\bibfnamefont {Yuta}\ \bibnamefont
  {Michimura}},\ }\bibfield  {title} {\enquote {\bibinfo {title} {{Optical Ring
  Cavity Search for Axion Dark Matter}},}\ }\href {\doibase
  10.1103/PhysRevLett.121.161301} {\bibfield  {journal} {\bibinfo  {journal}
  {Phys. Rev. Lett.}\ }\textbf {\bibinfo {volume} {121}},\ \bibinfo {pages}
  {161301} (\bibinfo {year} {2018})},\ \Eprint
  {http://arxiv.org/abs/1805.11753} {arXiv:1805.11753 [astro-ph.CO]}
  \BibitemShut {NoStop}%
\bibitem [{\citenamefont {Maggiore}(2007)}]{Maggiore:1900zz}%
  \BibitemOpen
  \bibfield  {author} {\bibinfo {author} {\bibfnamefont {Michele}\ \bibnamefont
  {Maggiore}},\ }\href {http://www.oup.com/uk/catalogue/?ci=9780198570745}
  {\emph {\bibinfo {title} {{Gravitational Waves. Vol. 1: Theory and
  Experiments}}}},\ Oxford Master Series in Physics\ (\bibinfo  {publisher}
  {Oxford University Press},\ \bibinfo {year} {2007})\BibitemShut {NoStop}%
\bibitem [{\citenamefont {Goryachev}\ \emph {et~al.}(2018)\citenamefont
  {Goryachev}, \citenamefont {McAllister},\ and\ \citenamefont
  {Tobar}}]{Goryachev:2018vjt}%
  \BibitemOpen
  \bibfield  {author} {\bibinfo {author} {\bibfnamefont {Maxim}\ \bibnamefont
  {Goryachev}}, \bibinfo {author} {\bibfnamefont {Ben}\ \bibnamefont
  {McAllister}}, \ and\ \bibinfo {author} {\bibfnamefont {Michael~E.}\
  \bibnamefont {Tobar}},\ }\bibfield  {title} {\enquote {\bibinfo {title}
  {{Axion Detection with Precision Frequency Metrology}},}\ }\href@noop {} {\
  (\bibinfo {year} {2018})},\ \Eprint {http://arxiv.org/abs/1806.07141}
  {arXiv:1806.07141 [physics.ins-det]} \BibitemShut {NoStop}%
\bibitem [{\citenamefont {Hecht}(2016)}]{hecht2016optics}%
  \BibitemOpen
  \bibfield  {author} {\bibinfo {author} {\bibfnamefont {E.}~\bibnamefont
  {Hecht}},\ }\href {https://books.google.com/books?id=Bv1RrgEACAAJ} {\emph
  {\bibinfo {title} {Optics}}}\ (\bibinfo  {publisher} {Pearson},\ \bibinfo
  {year} {2016})\BibitemShut {NoStop}%
\bibitem [{\citenamefont {Isogai}\ \emph {et~al.}(2013)\citenamefont {Isogai},
  \citenamefont {Miller}, \citenamefont {Kwee}, \citenamefont {Barsotti},\ and\
  \citenamefont {Evans}}]{Isogai:2013ic}%
  \BibitemOpen
  \bibfield  {author} {\bibinfo {author} {\bibfnamefont {T}~\bibnamefont
  {Isogai}}, \bibinfo {author} {\bibfnamefont {J}~\bibnamefont {Miller}},
  \bibinfo {author} {\bibfnamefont {P}~\bibnamefont {Kwee}}, \bibinfo {author}
  {\bibfnamefont {L}~\bibnamefont {Barsotti}}, \ and\ \bibinfo {author}
  {\bibfnamefont {M}~\bibnamefont {Evans}},\ }\bibfield  {title} {\enquote
  {\bibinfo {title} {{Loss in long-storage-time optical cavities}},}\
  }\href@noop {} {\bibfield  {journal} {\bibinfo  {journal} {Optics Express}\
  }\textbf {\bibinfo {volume} {21}},\ \bibinfo {pages} {30114} (\bibinfo {year}
  {2013})}\BibitemShut {NoStop}%
\bibitem [{\citenamefont {Bähre}\ \emph {et~al.}(2013)\citenamefont {Bähre}
  \emph {et~al.}}]{Bahre:2013ywa}%
  \BibitemOpen
  \bibfield  {author} {\bibinfo {author} {\bibfnamefont {Robin}\ \bibnamefont
  {Bähre}} \emph {et~al.},\ }\bibfield  {title} {\enquote {\bibinfo {title}
  {{Any light particle search II —Technical Design Report}},}\ }\href
  {\doibase 10.1088/1748-0221/8/09/T09001} {\bibfield  {journal} {\bibinfo
  {journal} {JINST}\ }\textbf {\bibinfo {volume} {8}},\ \bibinfo {pages}
  {T09001} (\bibinfo {year} {2013})},\ \Eprint {http://arxiv.org/abs/1302.5647}
  {arXiv:1302.5647 [physics.ins-det]} \BibitemShut {NoStop}%
\bibitem [{\citenamefont {Irastorza}\ \emph {et~al.}(2013)\citenamefont
  {Irastorza} \emph {et~al.}}]{Irastorza:1567109}%
  \BibitemOpen
  \bibfield  {author} {\bibinfo {author} {\bibfnamefont {Igor~G.}\ \bibnamefont
  {Irastorza}} \emph {et~al.} (\bibinfo {collaboration} {IAXO Collaboration}),\
  }\href {http://cds.cern.ch/record/1567109} {\emph {\bibinfo {title} {{The
  International Axion Observatory IAXO. Letter of Intent to the CERN SPS
  committee}}}},\ \bibinfo {type} {Tech. Rep.}\ \bibinfo {number}
  {CERN-SPSC-2013-022. SPSC-I-242}\ (\bibinfo  {institution} {CERN},\ \bibinfo
  {address} {Geneva},\ \bibinfo {year} {2013})\BibitemShut {NoStop}%
\bibitem [{\citenamefont {Budker}\ \emph {et~al.}(2014)\citenamefont {Budker},
  \citenamefont {Graham}, \citenamefont {Ledbetter}, \citenamefont
  {Rajendran},\ and\ \citenamefont {Sushkov}}]{Budker:2013hfa}%
  \BibitemOpen
  \bibfield  {author} {\bibinfo {author} {\bibfnamefont {Dmitry}\ \bibnamefont
  {Budker}}, \bibinfo {author} {\bibfnamefont {Peter~W.}\ \bibnamefont
  {Graham}}, \bibinfo {author} {\bibfnamefont {Micah}\ \bibnamefont
  {Ledbetter}}, \bibinfo {author} {\bibfnamefont {Surjeet}\ \bibnamefont
  {Rajendran}}, \ and\ \bibinfo {author} {\bibfnamefont {Alex}\ \bibnamefont
  {Sushkov}},\ }\bibfield  {title} {\enquote {\bibinfo {title} {{Proposal for a
  Cosmic Axion Spin Precession Experiment (CASPEr)}},}\ }\href {\doibase
  10.1103/PhysRevX.4.021030} {\bibfield  {journal} {\bibinfo  {journal} {Phys.
  Rev.}\ }\textbf {\bibinfo {volume} {X4}},\ \bibinfo {pages} {021030}
  (\bibinfo {year} {2014})},\ \Eprint {http://arxiv.org/abs/1306.6089}
  {arXiv:1306.6089 [hep-ph]} \BibitemShut {NoStop}%
\bibitem [{\citenamefont {Cumming}\ \emph {et~al.}(2012)\citenamefont {Cumming}
  \emph {et~al.}}]{Cook:2012tu}%
  \BibitemOpen
  \bibfield  {author} {\bibinfo {author} {\bibfnamefont {A.~V.}\ \bibnamefont
  {Cumming}} \emph {et~al.},\ }\bibfield  {title} {\enquote {\bibinfo {title}
  {{Design and development of the advanced LIGO monolithic fused silica
  suspension}},}\ }\href@noop {} {\bibfield  {journal} {\bibinfo  {journal}
  {Classical and Quantum Gravity}\ }\textbf {\bibinfo {volume} {29}},\ \bibinfo
  {pages} {035003} (\bibinfo {year} {2012})}\BibitemShut {NoStop}%
\bibitem [{\citenamefont {Gonzalez}\ and\ \citenamefont
  {Saulson}(1994)}]{Gonzalez1994}%
  \BibitemOpen
  \bibfield  {author} {\bibinfo {author} {\bibfnamefont {G}~\bibnamefont
  {Gonzalez}}\ and\ \bibinfo {author} {\bibfnamefont {P~R}\ \bibnamefont
  {Saulson}},\ }\bibfield  {title} {\enquote {\bibinfo {title} {{Brownian
  motion of a mass suspended by an anelastic wire}},}\ }\href@noop {}
  {\bibfield  {journal} {\bibinfo  {journal} {J. Acoust. Soc. Am.}\ }\textbf
  {\bibinfo {volume} {96}},\ \bibinfo {pages} {207--212} (\bibinfo {year}
  {1994})}\BibitemShut {NoStop}%
\bibitem [{\citenamefont {Abramovici}\ \emph {et~al.}(1996)\citenamefont
  {Abramovici} \emph {et~al.}}]{Abramovici:1996dz}%
  \BibitemOpen
  \bibfield  {author} {\bibinfo {author} {\bibfnamefont {A.}~\bibnamefont
  {Abramovici}} \emph {et~al.},\ }\bibfield  {title} {\enquote {\bibinfo
  {title} {{Improved sensitivity in a gravitational wave interferometer and
  implications for LIGO}},}\ }\href {\doibase 10.1016/0375-9601(96)00377-5}
  {\bibfield  {journal} {\bibinfo  {journal} {Phys. Lett.}\ }\textbf {\bibinfo
  {volume} {A218}},\ \bibinfo {pages} {157--163} (\bibinfo {year}
  {1996})}\BibitemShut {NoStop}%
\bibitem [{\citenamefont {Chou}\ \emph {et~al.}(2016)\citenamefont {Chou} \emph
  {et~al.}}]{Chou:2015sle}%
  \BibitemOpen
  \bibfield  {author} {\bibinfo {author} {\bibfnamefont {Aaron~S.}\
  \bibnamefont {Chou}} \emph {et~al.} (\bibinfo {collaboration} {Holometer}),\
  }\bibfield  {title} {\enquote {\bibinfo {title} {{First Measurements of High
  Frequency Cross-Spectra from a Pair of Large Michelson Interferometers}},}\
  }\href {\doibase 10.1103/PhysRevLett.117.111102} {\bibfield  {journal}
  {\bibinfo  {journal} {Phys. Rev. Lett.}\ }\textbf {\bibinfo {volume} {117}},\
  \bibinfo {pages} {111102} (\bibinfo {year} {2016})},\ \Eprint
  {http://arxiv.org/abs/1512.01216} {arXiv:1512.01216 [gr-qc]} \BibitemShut
  {NoStop}%
\bibitem [{\citenamefont {Aasi}\ \emph {et~al.}(2015)\citenamefont {Aasi} \emph
  {et~al.}}]{TheLIGOScientific:2014jea}%
  \BibitemOpen
  \bibfield  {author} {\bibinfo {author} {\bibfnamefont {J.}~\bibnamefont
  {Aasi}} \emph {et~al.} (\bibinfo {collaboration} {LIGO Scientific}),\
  }\bibfield  {title} {\enquote {\bibinfo {title} {{Advanced LIGO}},}\ }\href
  {\doibase 10.1088/0264-9381/32/7/074001} {\bibfield  {journal} {\bibinfo
  {journal} {Class. Quant. Grav.}\ }\textbf {\bibinfo {volume} {32}},\ \bibinfo
  {pages} {074001} (\bibinfo {year} {2015})},\ \Eprint
  {http://arxiv.org/abs/1411.4547} {arXiv:1411.4547 [gr-qc]} \BibitemShut
  {NoStop}%
\bibitem [{\citenamefont {Battesti}\ \emph {et~al.}(2018)\citenamefont
  {Battesti} \emph {et~al.}}]{Battesti:2018bgc}%
  \BibitemOpen
  \bibfield  {author} {\bibinfo {author} {\bibfnamefont {Rémy}\ \bibnamefont
  {Battesti}} \emph {et~al.},\ }\bibfield  {title} {\enquote {\bibinfo {title}
  {{High magnetic fields for fundamental physics}},}\ }\href {\doibase
  10.1016/j.physrep.2018.07.005} {\bibfield  {journal} {\bibinfo  {journal}
  {Phys. Rept.}\ }\textbf {\bibinfo {volume} {765-766}},\ \bibinfo {pages}
  {1--39} (\bibinfo {year} {2018})},\ \Eprint {http://arxiv.org/abs/1803.07547}
  {arXiv:1803.07547 [physics.ins-det]} \BibitemShut {NoStop}%
\end{thebibliography}%
\end{document}